\documentclass{article} 
\usepackage{iclr2025_conference,times}

\usepackage{amsmath,amsfonts,bm}

\def\eqref#1{equation~\ref{#1}}

\def\1{\bm{1}}

\DeclareMathAlphabet{\mathsfit}{\encodingdefault}{\sfdefault}{m}{sl}
\SetMathAlphabet{\mathsfit}{bold}{\encodingdefault}{\sfdefault}{bx}{n}

\usepackage{hyperref}
\usepackage{url}
\usepackage{multirow}
\usepackage{bigfoot}
\usepackage{adjustbox}
\usepackage{booktabs}
\usepackage{graphicx}   
\usepackage{subcaption} 
\usepackage{caption}
\usepackage[colorinlistoftodos, textwidth=3cm]{todonotes}
\usepackage{comment}
\usepackage{float}
\usepackage{booktabs}
\usepackage[table]{xcolor}

\title{Granite Embedding Multilingual R2 Models}

\author{Granite Embedding Team, 
IBM Research\thanks{See Section \ref{sec:contributions} for full author list. For questions, comments, compliments contact awasthyp@us.ibm.com or raduf@us.ibm.com.} \thanks{For feedback or comments on this work, please open an issue at \url{https://github.com/ibm-granite/granite-embedding-models}.}
}

\iclrfinalcopy 
\begin{document}

\maketitle

\begin{abstract}

We introduce the multilingual Granite Embedding R2 models, a family of encoder-based embedding models for enterprise-scale dense retrieval across 200+ languages. Extending our English-focused R2 release, these models add enhanced support for 52 languages and programming code, a 32,768-token context window (a 64x expansion over R1), and state-of-the-art overall performance across multilingual and cross-lingual text search, code retrieval, long-document search, and reasoning retrieval datasets. The release consists of two bi-encoder models based on the ModernBERT architecture with an expanded multilingual vocabulary: a 311M-parameter full-size, and a 97M-parameter compact model built via model pruning and vocabulary selection that achieves the highest retrieval score of any open multilingual embedding model under 100M parameters. The full-size also supports Matryoshka Representation Learning for flexible embedding dimensionality. Both models are trained on enterprise-appropriate data with governance oversight, and released under the Apache 2.0 license at \url{https://huggingface.co/collections/ibm-granite}, designed to support responsible use and enable unrestricted research and enterprise adoption.

\end{abstract}

\begin{figure}[h]
    \centering
    \makebox[\textwidth][c]{%
        \includegraphics[width=1.1\textwidth]{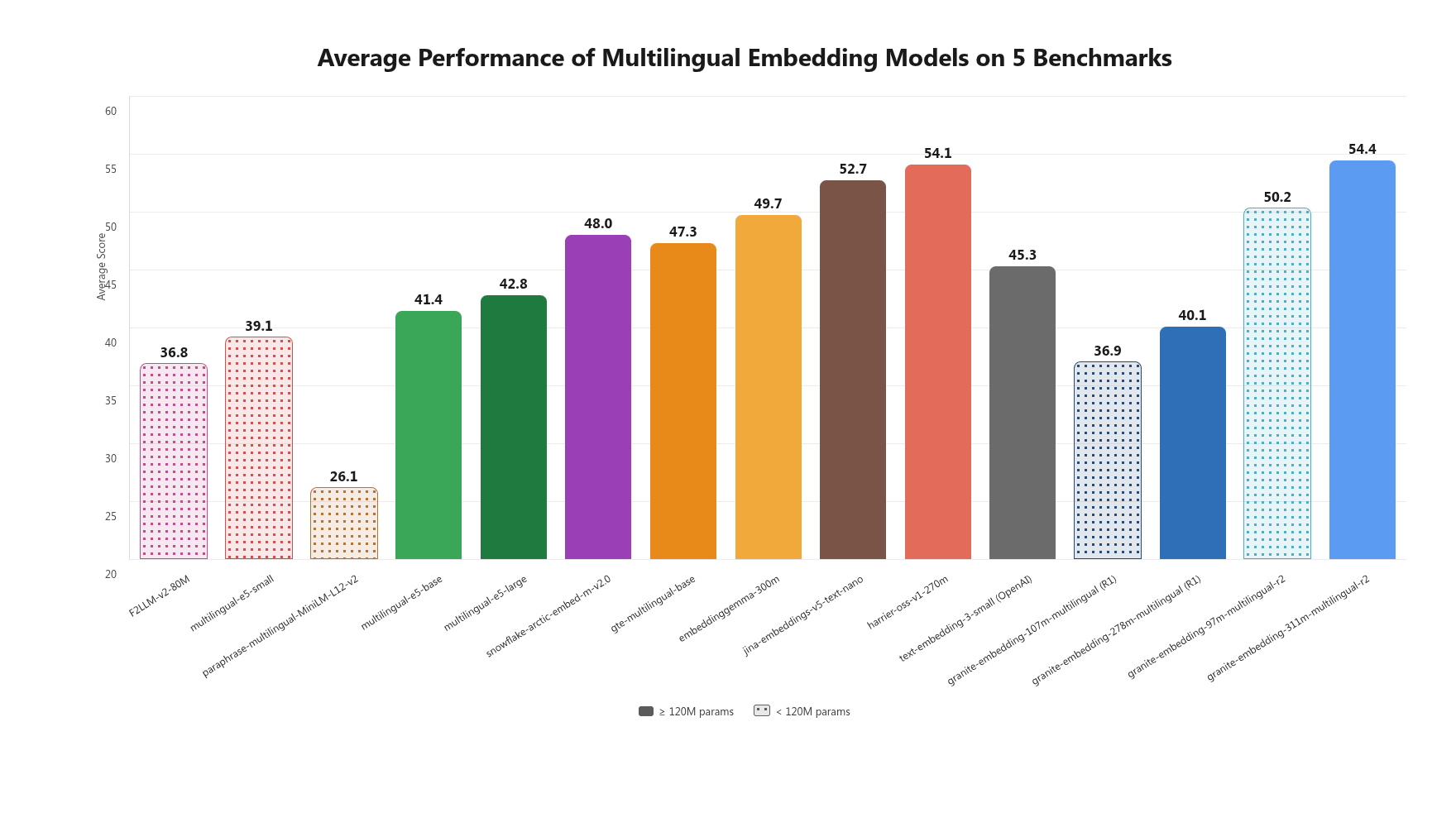}%
    }
    \caption{Performance of multilingual Granite R2 embedding models and comparable size open-source models, average multilingual MTEB performance. Refer section \ref{subsec:retriever_performance} for details. }
    \label{fig:avg-perf}
\end{figure}

\section{Introduction}

Bi-encoder text embedding models map text to a fixed-dimension vector such that semantically similar texts lie close in the vector space. These embeddings are most commonly used in retrieval, where document relevance to a query is scored by embedding similarity \citep{dunn2017searchqa, xiong2020approximatenearestneighbornegative, neelakantan2022textcodeembeddingscontrastive, 10.1145/3269206.3271800, zhao2020spartaefficientopendomainquestion}, as well as in document clustering \citep{angelov20} and text classification \citep{sun2019fine}.

Encoder-based embedding models \citep{wang2022e5, bge_embedding, chen2024bgem3, merrick2024arcticembedv1, zhang2024mgte, wang2024multilingual, nussbaum2024nomic, awasthy2025graniteembeddingr2models} are widely used for retrieval due to their low inference latency and small memory footprint compared to decoder-based models \citep{lee2024nvembed, wang2023e5mistral,qwen3embedding, akram2026jina}. Building effective multilingual embedding models, however, poses additional challenges: the vocabulary must capture morphological diversity across many language families, the encoder must learn aligned representations across languages and scripts, and training data must cover a broad set of languages while maintaining quality. Many existing multilingual models are trained on data with non-commercial licenses, have limited context length, or provide uneven coverage across languages.

This report introduces the Granite Embedding Multilingual R2 models, purpose-built for multilingual information retrieval. These models provide substantial improvements over the R1 multilingual models~\citep{awasthy2025granite}: a 64x expanded context length (from 512 to 32,768 tokens), support for 200+ languages with enhanced support for 52 languages and programming code, and an updated encoder based on the ModernBERT architecture \citep{warner2024modernbert} with a multilingual tokenizer trained on text and code across 200+ languages. Training data is curated and screened to remove personal information and profane language. Notably, we exclude the popular MS-MARCO dataset due to its non-commercial license. The models are released under the Apache 2.0 license. We provide two sizes to accommodate different inference budgets:
\begin{itemize}
    \item \texttt{granite-embedding-311m-multilingual-r2} (311M parameters)\footnote{\href{https://huggingface.co/ibm-granite/granite-embedding-311m-multilingual-r2}{ibm-granite/granite-embedding-311m-multilingual-r2}}: the full-size model, with a 768-dimensional output embedding and Matryoshka Representation Learning support. Replaces granite-embedding-278m-multilingual.
    \item \texttt{granite-embedding-97m-multilingual-r2} (97M parameters)\footnote{\href{https://huggingface.co/ibm-granite/granite-embedding-97m-multilingual-r2}{ibm-granite/granite-embedding-97m-multilingual-r2}}: a compact model built via distillation and vocabulary selection, with a 384-dimensional output embedding. Replaces granite-embedding-107m-multilingual.
\end{itemize}

Both models deliver state-of-the-art overall performance across multilingual retrieval (MTEB-v2 Retrieval, BEIR), code retrieval (COIR), long-document search (LongEmbed), cross-lingual retrieval (MLQA), and reasoning retrieval (BRIGHT, RAR-b), while supporting 32,768-token contexts. The 311M full-size scores 65.2 on MTEB-v2 Retrieval Avg, placing it in the top 3 of open multilingual embedding models under 500M parameters. The 97M compact model scores 60.3 --- the highest of any open multilingual embedding model under 100M parameters, with a 9+ point lead over the next-best model in its class.

The remainder of the paper is organized as follows: Section \ref{sec:encoder} describes training of the multilingual ModernBERT encoder. Section \ref{sec:biencoder} details the bi-encoder training recipes, including the vocabulary selection approach for the compact model. Section \ref{sec:evaluation} evaluates the Granite Embedding Multilingual models against other open-source multilingual encoder embedding models.

\section{Granite Multilingual Encoder Models}
\label{sec:encoder}

The Granite Embedding Multilingual R2 models feature updated encoder models, with longer context lengths, a richer multilingual training corpus, and modern architectural improvements. In this section, we discuss the architecture, training recipe and details of the high-quality multilingual corpus used to train the Granite Multilingual Encoder models. These models form the underlying backbone of the Granite Embedding models.

\subsection{Encoder Model Architecture}

The Granite Multilingual Encoder models have been trained following the mmbert training recipe \citep{marone2025mmbertmodernmultilingualencoder}, including modern model optimizations introduced by ModernBERT \citep{warner2024modernbert} such as alternating attention mechanism, rotary positional embeddings for flexible context length, and streamlined parameters (eliminating unnecessary bias terms). The models also support Flash Attention \citep{dao2023flashattention2} for improved efficiency. A key distinction from the English R2 encoder models is the use of an expanded multilingual tokenizer trained on code and text data across 200+ languages, with a vocabulary size of 262,152 for the base model and 180,000 for the compact model (via vocabulary selection). These encoder models have been trained on multilingual text and code data.

We train two models:
\begin{itemize}
    \item \emph{granite-encoder-multilingual} (311M parameters): the backbone of granite-embedding-311m-multilingual-r2. Its architecture follows ModernBERT-base, with 22 layers, a vector size of 768, and GeLU activation. Similar to ModernBERT, the model uses alternating global attention in every third layer.
    \item \emph{granite-encoder-small-multilingual} (97M parameters): the backbone of granite-embedding-97m-multilingual-r2. Built via model pruning and vocabulary selection from the larger model, with 12 layers, a vector size of 384, SiLU activation, and a compact 180K-token vocabulary that preserves broad multilingual coverage while reducing model size by approximately 3x.
\end{itemize}

Detailed specifications of the architecture of each model are shown in Table~\ref{tab:architecture}.

\begin{table}[!t]
    \centering
    \setlength{\tabcolsep}{0.5em} 
    {\renewcommand{\arraystretch}{1.3}
    \begin{tabular}{l|cc}
    \hline
         & granite-encoder-small-multilingual & granite-encoder-multilingual\\ \hline
        Embedding size & $384$ & $768$ \\
        Layers & $12$ & $22$  \\
        Attention heads & $12$ & $12$  \\
        Intermediate size & $1536$ & $1152$  \\
        Activation Function & SiLU & GeLU \\
        Vocabulary Size & $180{,}000$  & $262{,}152$ \\
        Max. Seq. Length Trained & $8{,}192$ & $8{,}192$ \\
        Global Rope Theta & $160{,}000$ & $160{,}000$  \\
        Parameters & $\sim$97M & $\sim$311M \\
        Active Params & $\sim$28M & $\sim$110M \\
        \hline
    \end{tabular}}
    \caption{Architectural Details for Granite Multilingual Encoder Models}
    \label{tab:architecture}
\end{table}

The underlying encoder was pretrained on text from 200+ languages, and we report general-purpose embeddings for any of them. In addition, we provide enhanced support for 52 languages and programming code that receive explicit retrieval-pair and cross-lingual training data, producing higher-quality embeddings on retrieval tasks. The supported languages are listed in Appendix ~\ref{app:languages}.

\subsection{Tokenizer Fertility Analysis}
\label{sec:tokenizer-fertility}

Tokenizer efficiency directly affects both throughput and effective context length: a model with a 32K-token window but twice the fertility of a competitor can effectively encode only half as much text. Tables \ref{tab:fertility} and \ref{tab:fertility-code} report tokenizer fertility (tokens per word) across natural languages and programming languages, respectively. Our full-size model uses the \texttt{Gemma3} tokenizer \citep{gemmateam2025gemma3technicalreport} with a 262K-token vocabulary  (referred to as \texttt{granite-multi-262K}), while the compact variant uses a customized \texttt{gpt-oss} tokenizer pruned to 180K tokens (\texttt{granite-multi-180K}). On multilingual Wikipedia text, the Granite tokenizers exhibit 10\% higher fertility than \texttt{XLM‑RoBERTa}–based tokenizers (used by popular open source embedding models like bge‑m3 \citep{chen2024bgem3}, multilingual‑e5 \citep{wang2024multilinguale5}, and gte‑multilingual \citep{zhang2024mgte}). However, on programming languages, both Granite tokenizers consistently achieve lower or comparable fertility relative to \texttt{XLM‑RoBERTa} tokenizer, indicating more efficient encoding of code data. We additionally include the unpruned \texttt{gpt-oss} tokenizer to show that pruning has a negligible effect on fertility.

\begin{table}[!ht]
    \centering
    {\renewcommand{\arraystretch}{1.3}
    \begin{adjustbox}{width=0.9\textwidth}
    \begin{tabular}{l|c|cccccccccc|c}
    \toprule
        Tokenizer & en & ar & de & es & hi & ja & ko & ru & sw & te & zh & Avg. \\
        \hline
        Granite-multi-262K
        & 1.60 & 2.12 & 2.00 & 1.56 & 1.52 & 0.68 & 0.79 & 2.24 & 2.12 & 3.02 & 0.77 & 1.67 \\
        Granite-multi-180K
        & 1.50 & 1.96 & 1.92 & 1.52 & 1.81 & 0.82 & 0.81 & 2.25 & 1.89 & 3.24 & 0.86 & 1.69 \\
        XLM-R-based, 250K
        & 1.67 & 1.83 & 1.84 & 1.53 & 1.61 & 0.68 & 0.72 & 2.04 & 1.69 & 2.41 & 0.74 & 1.52 \\
        GPT-OSS
        & 1.50 & 1.96 & 1.92 & 1.52 & 1.81 & 0.82 & 0.81 & 2.25 & 1.89 & 3.24 & 0.86 & 1.69 \\
    \bottomrule
    \end{tabular}
    \end{adjustbox}}
    \caption{Tokenizer fertility (tokens per word) across natural languages. Lower values indicate more efficient encoding — fewer tokens per word allow more text to fit within a fixed context window. Measured on a sample of 10K sentences per language from Wikipedia.}
\label{tab:fertility}
\end{table}

\begin{table}[!ht]
    \centering
    {\renewcommand{\arraystretch}{1.3}
    \begin{adjustbox}{width=0.9\textwidth}
    \begin{tabular}{l|cccccccc|c}
    \toprule
        Tokenizer & Python & Java & JavaScript & PHP & Ruby & SQL & C & C++ & Avg. \\
        \hline
        Granite-base-262K & 3.46 & 3.04 & 3.82 & 3.41 & 3.52 & 3.49 & 3.18 & 3.38 & 3.41 \\
        Granite-small-180K & 2.81 & 2.49 & 3.13 & 2.93 & 2.96 & 2.83 & 2.54 & 2.74 & 2.80 \\
        XLM-R-based, 250K & 3.73 & 3.32 & 4.16 & 3.78 & 3.55 & 3.58 & 3.30 & 3.62 & 3.63 \\
        GPT-OSS & 2.81 & 2.48 & 3.12 & 2.92 & 2.96 & 2.83 & 2.53 & 2.74 & 2.80 \\
    \bottomrule
    \end{tabular}
    \end{adjustbox}}
    \caption{Tokenizer fertility (tokens per word) across programming languages. Lower values indicate more efficient encoding — fewer tokens per word allow more text to fit within a fixed context window. Measured on a sample of 10K sentences per language from github-code, except for SQL which uses around 2.8K due to limited data availability.}
    \label{tab:fertility-code}
\end{table}

\subsection{Training Data}
\label{sec:encoder-data}

We curate a diverse, high-quality multilingual corpus of text and code to train our encoder models. The largest component of the training data is FineWeb2 \citep{penedo2025fineweb2pipelinescale}, from which we filter and retain data covering more than 1{,}800 languages throughout training. English-language data is primarily drawn from GneissWeb \citep{gohari2025gneissweb}, an IBM-curated dataset composed exclusively of open, commercial-friendly sources and rigorously filtered to produce high-quality corpora for language model training. For the last two stages of encoder training, which require higher-quality datasets, we use the filtered FineWeb-Edu \citep{lozhkov2024fineweb-edu} for multilingual data. To enhance domain and stylistic diversity, we additionally include multilingual Wikipedia, Stack Exchange, and arXiv data. For improved performance on code-related tasks, we incorporate a subset of code data from the training corpora of the Granite Code models \citep{mishra2024granite}. 

We follow the language sampling strategy of mmBERT \citep{marone2025mmbertmodernmultilingualencoder} to balance coverage between high-resource and low-resource languages throughout training. Specifically, we progressively increase the number of covered languages while decreasing the temperature used for language sampling from Stage 1 to Stage 3 (as described in ~\ref{sec:encoder-training}), thereby promoting broader multilingual exposure in later stages. As a result, the final training stage covers more than 1,800 languages.

For tokenization, we use the Gemma3 tokenizer \citep{gemmateam2025gemma3technicalreport} with a 262K-token vocabulary for the base model. The compact model employs a customized GPT-OSS tokenizer, further pruned to a 180K-token vocabulary, which preserves broad multilingual coverage while enabling a smaller model footprint.

\subsection{Training Recipe}
\label{sec:encoder-training}

Following the ModernBERT training setting and mmBERT \citep{marone2025mmbertmodernmultilingualencoder} recipe, we train our base encoder model on the Masked Language Modeling objective over three distinct stages:

\begin{enumerate}
    \item Large Scale Pretraining: First, we train on 2.5 trillion tokens of multilingual text data, with a maximum context length of 1024. We use a Warmup-Stable-Decay learning rate schedule \citep{hu2024minicpm}, with a RoPE theta of 10,000.
    \item Context Extension: We then scale up the context length to 8192 and the RoPE theta to 160,000, continuing training on an additional 600 billion tokens at a constant learning rate.
    \item Learning Rate Decay: Finally, we train with the same context length and RoPE theta as in Stage 2, but with a \texttt{1-sqrt} learning rate decay from the peak learning rate for 100 billion tokens. Following the mmBERT approach \citep{marone2025mmbertmodernmultilingualencoder}, we train three variants during this stage and linear merge them to form the final model: 
    \begin{enumerate}
        \item an English-focused variant with higher proportions of English data
        \item a multilingual-focused variant trained on data spanning more than 1,800 languages
        \item a variant continued directly from Stage 2
    \end{enumerate}
\end{enumerate}

Across all stages, we use the StableAdamW optimizer \citep{wortsman2023stable}, and employ efficient training mechanisms such as sequence packing, unpadding, and flash attention, as described in \cite{warner2024modernbert}.

The small model follows the same three-stage training recipe as the base model, with several adjustments. We initialize its weights from \texttt{granite-embedding-small-english-r2} \citep{awasthy2025graniteembeddingr2models} by directly reusing the embedding rows for tokens shared with the source vocabulary and taking the average of all source embeddings for newly introduced tokens. This approach preserves learned representations for shared tokens while providing a reasonable initialization for new ones, enabling faster convergence during continued pretraining. In addition, during Stage 1, we reduce the masking rate and halve the learning rate and weight decay after 949 billion tokens. Finally, Stage 3 is extended to 200 billion tokens, compared to 100 billion tokens in the base model.

\section{Granite Embedding Multilingual R2}
\label{sec:biencoder}

The Granite Embedding Multilingual R2 models are purpose-built for retrieval and trained on carefully curated, enterprise-ready data. This section describes the training data and methodology, covering contrastive finetuning, knowledge distillation, model merging, and pruning with vocabulary selection.

\subsection{Training Data}
\label{sec:retriever-data}

Granite Embedding Multilingual Models are trained on four types of data:
\begin{enumerate}
    \item Unsupervised title-body paired data scraped from the web
    \item Publicly available paired data with permissive, enterprise-friendly licenses
    \item IBM product documentation paired data targeting specific technical domains
    \item IBM-generated synthetic data, including multilingual long-document and short-passage data, and reasoning-oriented data
\end{enumerate}

All data undergoes a clearance process with technical, business, and governance review, capturing content description, intended use, data classification, licensing, usage restrictions, and assessment of sensitive information (e.g., personal information).

The Multilingual R2 models reuse and extend the English Granite Embedding R2 training data \citep{awasthy2025granite}, adding multilingual and cross-lingual pairs:
\begin{itemize}
    \item Multilingual Retrieval Data: retrieval pairs across the 52 enhanced-support languages, drawn from publicly available multilingual datasets and synthetically generated long- and short-passage pairs.
    \item Cross-lingual Data: pairs across multiple language combinations to enable cross-lingual retrieval (e.g., querying in one language and retrieving in another), including parallel translations of the same text.
    \item Code Data: code retrieval pairs from diverse sources, with hard negatives mined in most cases. Covers Python, Go, Java, JavaScript, PHP, Ruby, SQL, C++, and C.
    \item Multi-Turn Conversational IR Data: multi-turn data for conversational retrieval.
    \item Synthetic Multilingual Data: a synthetic data generation (SDG) pipeline producing passage-retrieval training data across 18 languages. Passages sampled from MIRACL corpora (where available) or Wikipedia serve as conditioning context for gpt-oss-120b, which generates question-answer pairs via few-shot prompting. Each query is augmented with 10 BM25 hard negatives, and data is produced at both passage and document granularity. We apply two rounds of LLM-as-judge filtering: removing false negatives from contrastive pairs, and scoring queries for human-likeness using language-specific prompts.
    \item Synthetic Reasoning Data: reasoning-oriented queries generated with gpt-oss-120b for documents from Arxiv, Pubmed, StackExchange Math, and StackOverflow, with hard negatives mined or generated following the method of \citet{shao2025reasonirtrainingretrieversreasoning}.
\end{itemize}

\subsection{Training Recipe}
\label{subsec: general training recipe}

Embedding models are typically trained with a contrastive objective \citep{gao-etal-2021-simcse} that pulls query embeddings toward relevant passages and pushes them away from non-relevant ones. The Granite Embedding Multilingual models are trained with the following pipeline:

\begin{enumerate}
    \item Contrastive Finetuning: the models are first finetuned on a large corpus of multilingual paired data using the improved contrastive loss from \citet{li2023gte}. For a batch $([q_i,(p_{ij})_j])_i$ of queries and passages --- where $p_{i0}$ is the positive passage for query $i$ and $p_{ij}, j>0$ are negatives --- the loss is:

    \[
        \mathcal{L}_{C} = -\frac{1}{n} \sum_{i=1}^{n} \mathrm{log} \frac{e^{s(q_i, p_{i0})}}{Z_i}
    \label{eq:cont-loss}
    \]
    \[
    \begin{aligned}
    Z_i = e^{s(q_i, p_{i0})} + \alpha\sum_{j>0} e^{s(q_i,p_{ij})} + \beta\sum_{i' \neq i} e^{s(q_i,q_{i'})} + \gamma\sum_{j>0} e^{s(p_{i0},p_{ij})}
    \label{eq:z-comp}
    \end{aligned}
    \]

    and $s(q, p)$ is the temperature-scaled cosine similarity between the \texttt{[CLS]} embeddings of $q$ and $p$:
    \[
        s(q,p) = \frac{1}{\tau} \frac{\mathbf{E}(q)_\texttt{[CLS]} \cdot\mathbf{E}(p)_\texttt{[CLS]}} {\|\mathbf{E}(q)_\texttt{[CLS]}\|\|\mathbf{E}(p)_\texttt{[CLS]}\|}
    \label{eq:sim-score}
    \]

    We use a large batch size with in-batch negatives to better approximate the contrastive objective.

    \item Knowledge Distillation: We distill from multiple teacher models into the student, minimizing the cross entropy between the teacher's similarity-score distribution $Sim_t$ and the student's $Sim_s$. Following \citet{hinton2014distilling}, both distributions are scaled by a temperature $\tau_{KD}$:
    \[
    Sim_\ast(q_i, p_{ij}) =
    \frac{\exp\!\left(s_\ast(q_i, p_{ij})/\tau_{KD}\right)}
         {\sum_{k=1}^{m}\exp\!\left(s_\ast(q_i, p_{ik})/\tau_{KD}\right)},
    \qquad \ast \in \{t, s\},
    \]
    and minimize
    \[
        \label{eq:kd-loss}
        \mathcal{L}_{KD} = - \sum_{i=1}^{n} \sum_{j=1}^{m}
            Sim_t(q_i, p_{ij}) \, \log Sim_s(q_i, p_{ij}).
    \]
    We use mined hard negatives and two teachers: one specialized on English retrieval, and one with stronger multilingual capabilities. For each homogeneous batch, the teacher is selected according to the language of the data.

    \item Context Extension: Knowledge distillation is performed in two phases. The first uses a maximum sequence length of 512 with a large effective batch size; the second extends the maximum length to 4k and increases the RoPE theta to improve long-context performance.

    \item Model Merging: To improve English performance, we train an identical model on English-only data and merge its weights with the multilingual model.

    \item Matryoshka Representation Learning (311M only): The full-size model is trained with Matryoshka Representation Learning \citep{kusupati2024matryoshkarepresentationlearning}, allowing 768-dimensional embeddings to be truncated to 512, 384, 256, or 128 dimensions with graceful degradation, reducing storage, memory, and latency costs.
\end{enumerate}

\subsection{Training Stage Ablation}
\label{subsec:stage-ablation}

\begin{table}[!t]
    \centering
    {\renewcommand{\arraystretch}{1.3}
    \begin{tabular}{l|c|c|c}
    \toprule
        Training Stage (cumulative) & ML MTEB-v2 Retrieval & MIRACL & MTEB Code  \\
        \hline
        Granite Encoder (base)      & 45.0 & 43.5 & 36.4   \\
        + Contrastive FT            & 55.8 & 49.6  & 53.5    \\
        + Knowledge Distillation    &  61.2 &  60.0 & 62.0   \\
        + Context Extension         &  63.7 &  60.1 & 63.2   \\
        + Model Merging             & 65.2 & 59.8 & 63.9    \\
    \bottomrule
    \end{tabular}}
    \caption{Cumulative effect of each training stage on granite-embedding-311m-multilingual-r2 model performance. Each row adds one stage to all previous stages, showing the marginal contribution of each step. We reoprt the average NDCG@10 performance across all tasks of a benchmarl }
\label{tab:stage-ablation}
\end{table}

To validate the contribution of each pipeline stage, we evaluate the checkpoint after each stage on a representative set of benchmarks. Table \ref{tab:stage-ablation} shows the cumulative effect of each stage on granite-embedding-311m-multilingual-r2. For the encoder, we report performance after finetuning for a few hundred steps on Miracl triples on the contrastive objective.

\subsection{Compact Model: Vocabulary Selection}
\label{subsec:compact-model}

The granite-embedding-97m-multilingual-r2 model uses a pruned multilingual tokenizer with 180K vocabulary tokens based on the GPT-OSS vocabulary, reduced from 200K by removing the most infrequent tokens, preserving broad coverage across 200+ languages at a smaller footprint.

The model is then finetuned with knowledge distillation and contrastive training. Despite being approximately 3x smaller than the full-size, it scores 60.3 on MTEB-v2 Retrieval Avg --- the highest of any open multilingual embedding model under 100M parameters.

\subsection{Teacher Training}
\label{subsec:teacher}

We employ larger decoder models to be used as teachers, finetuning them on the contrastive learning objective to produce strong embeddings. Following the approach for teachers used in the English R2 models \citep{awasthy2025granite}, we merge high-performing checkpoints (selected with a held-out vaildation set), with either spherical interpolation or equal-weight linear interpolation using mergekit \citep{goddard-etal-2024-arcees}, to create the final teacher models. After vigorous experiments with different models (from the Granite \citep{granite_4.1_llms}, Mistral \citep{jiang2023mistral7b}, Ministral \citep{liu2026ministral3} and Phi \citep{abdin2024phi4technicalreport} families), pooling strategies, hard negatives, instruction usage, and attention mechanism, we select two teachers for distilling the Granite Multilingual R2 embedding models:
\begin{itemize}
    \item English Teacher: we train a Mistral 7B\footnote{mistralai/Mistral-7B-Instruct-v0.2} teacher \citep{jiang2023mistral7b} on only English data. This is the same teacher used in the English R2 models\citep{awasthy2025graniteembeddingr2models}.
    \item Multilingual Teachers: We use the multilingual data in Section~\ref{sec:retriever-data} to train Granite 3.3 8B\footnote{ibm-granite/granite-3.3-8b-instruct} and Granite 4.1 8B\footnote{ibm-granite/granite-4.1-8b} teachers, finding the former teacher to be best for the 311M parameter embedding model and the latter to be better for the 97M embedding model.
\end{itemize}

For all teachers, we find last-token pooling using \texttt{[EOS]} to be an effective pooling mechanism for auto-regressive decoders, as shown by others \citep{qwen3embedding, akram2026jina}. While the Mistral teacher is adapted to use bi-directional attention, we find that the Granite-based decoder teachers show strong performance out-of-the box with causal attention. We also find that using a separate teacher for English and Multilingual data helps improve overall performance, especially for the smaller model.

\section{Evaluation}
\label{sec:evaluation}

We evaluate the performance of our multilingual models on a variety of tasks and domains, spanning multilingual retrieval, code retrieval, long-document search, and reasoning retrieval.  

\subsection{Retrieval Performance}
\label{subsec:retriever_performance}

We evaluate the Granite Embedding Multilingual models on a variety of retrieval tasks, spanning multiple domains, languages, document lengths and text objects:
\begin{itemize}
    \item Multilingual Retrieval: We evaluate on the MTEB-v2 Retrieval benchmark \citep{enevoldsen2025mmtebmassivemultilingualtext} 
    \item Code Retrieval: We evaluate on code retrieval tasks of the MTEB-Code benchmark \citep{enevoldsen2025mmtebmassivemultilingualtext}, including tasks from COIR \citep{li2024coircomprehensivebenchmarkcode}, which consists of text-to-code, code-to-text, and hybrid code retrieval.
    \item English Retreival: We evaluate on general information retrieval benchmarks such as  \citep{enevoldsen2025mmtebmassivemultilingualtext}, comprising retrieval tasks on a variety of domains with a focus on zero-shot evaluations.
    \item Long Context Retrieval: To evaluate performance on retrieving long-context documents, we measure performance on the LongEmbed benchmark \citep{zhu2024longembed}.
    \item Reasoning Retrieval: To evaluate reasoning retrieval, we measure performance on the Reasoning-as-Retrieval benchmark \citep{xiao2024rar}.
\end{itemize}

We compare our models with other state-of-the-art multilingual embedding models, including \texttt{multilingual-e5-base} \citep{wang2022e5}, \texttt{multilingual-e5-small}, \texttt{gte-multilingual-base} \citep{zhang2024mgte}, \texttt{snowflake-arctic-embed-m-v2.0} \citep{yu2024arcticembedv2}, \texttt{embeddinggemma-300m} \citep{vera2025embeddinggemmapowerfullightweighttext}, \texttt{jina-embeddings-v5-text-nano} \citep{akram2026jina}, and \texttt{harrier-oss-v1-270m}. We also compare to the R1 Granite Embedding Multilingual Models, to quantify the improvement over the previous release.

\subsubsection{Multilingual Retrieval Performance}

\begin{table}[!t]
    \centering
    \begin{adjustbox}{width=1.0\textwidth}
    {\renewcommand{\arraystretch}{1.3}
    \begin{tabular}{l|c|c|c|c|c|c|c}
\toprule
Model & Params & Embed. & MTEB Retr. & MTEB Code & MTEB Retr. & LongEmbed & RaR-b \\
 & (M) & Size & Multi (18) & (12) & En (10) & (6) & (17) \\
\hline
f2llm-v2-80m & 80 & 320 & 50.1 & 68.0 & 47.5 & 31.7 & 17.9 \\
multilingual-e5-small & 96 & 384 & 50.9 & 51.3 & 46.5 & 38.8 & 20.3 \\
multilingual-e5-base & 278 & 768 & 52.7 & 52.6 & 49.0 & 40.5 & 23.4 \\
snowflake-arctic-embed-m-v2.0 & 305 & 768 & 54.8 & 55.2 & 58.4 & 55.4 & 23.3 \\
gte-multilingual-base & 305 & 768 & 57.2 & 57.5 & 50.8 & 62.1 & 19.0 \\
embeddinggemma-300m & 300 & 768 & 62.5 & 69.0 & 54.6 & 55.4 & 26.1 \\
jina-embeddings-v5-text-nano & 239 & 768 & 63.3 & 71.2 & 58.8 & 63.6 & 25.2 \\
harrier-oss-v1-270m & 270 & 640 & 66.4 & 62.4 & 52.1 & 65.0 & 32.9 \\
\midrule
granite-embedding-107m-multilingual & 107 & 384 & 48.1 & 47.9 & 40.7 & 34.3 & 17.1 \\
granite-embedding-278m-multilingual & 278 & 768 & 52.2 & 48.5 & 51.5 & 37.7 & 18.9 \\
\textbf{granite-embedding-97m-multilingual-r2} & \textbf{97} & \textbf{384} & \textbf{60.3} & \textbf{60.5} & \textbf{50.1} & \textbf{62.9} & \textbf{24.9} \\
\textbf{granite-embedding-311m-multilingual-r2} & \textbf{311} & \textbf{768} & \textbf{65.2} & \textbf{63.9} & \textbf{52.6} & \textbf{71.7} & \textbf{28.0} \\
\bottomrule
\end{tabular}}
    \end{adjustbox}
    \caption{Multilingual Retrieval Performance. Scores are averages across tasks, with the number of tasks indicated in parentheses. All scores are NDCG@10 unless otherwise noted. }
\label{tab:performance}
\end{table}

We evaluate multilingual performance of our models on a variety of open benchmarks, spanning across multiple domains. For competitor models, we take results from the MTEB leaderboard when available. For MTEB Retrieval (English and Multilingual), we use a maximum sequence length of 1024, for Code and RaR-b we use an MSL of 8192, and for LongEmbed we use an MSL of 32K (for models with shorter context lengths, we truncate to the model's max seq. length.)

As shown in Table \ref{tab:performance}, the Granite Embedding Multilingual R2 models demonstrate strong performance across diverse multilingual tasks. The full-size granite-embedding-311m-multilingual-r2 scores 65.2 on MTEB-v2 Retrieval Avg, a +13 point improvement over granite-embedding-278m-multilingual (52.2), placing it in the top 3 of open multilingual embedding models under 500M parameters on MTEB-v2 Retrieval Avg. The compact granite-embedding-97m-multilingual-r2, at just 97M parameters, scores 60.3 --- a 9+ point lead over the next-best open model under 100M parameters.

\subsubsection{English-Only Performance: Multilingual vs.\ English R2}

\begin{table}[!t]
    \centering
    \begin{adjustbox}{width=1.0\textwidth}
    {\renewcommand{\arraystretch}{1.3}
    \begin{tabular}{l|c|c|c|c|c}
    \toprule
        Model                               &  Params  &  Embed.  &  MTEB-v2 &  MTEB Code &  LongEmbed  \\
        & (M) & Size & Retrieval (10)  &  (12)  & (6)  \\
        \hline
        granite-embedding-small-english-r2 & 47 & 384 & 53.9 & 55.8 & 63.7 \\
granite-embedding-english-r2 & 149 & 768 & 56.4 & 57.2 & 65.6 \\
\midrule
\textbf{granite-embedding-97m-multilingual-r2} & \textbf{97} & \textbf{384} & \textbf{50.1} & \textbf{60.5} & \textbf{65.5} \\
\textbf{granite-embedding-311m-multilingual-r2} & \textbf{311} & \textbf{768} & \textbf{52.6} & \textbf{63.9} & \textbf{71.7} \\
    \bottomrule
    \end{tabular}}
    \end{adjustbox}
    \caption{English-only retrieval performance comparison between the Granite English R2 and Multilingual R2 models, evaluated on the same English benchmarks. This quantifies the cost (if any) of multilingual capability on English tasks.}
\label{tab:english-vs-multilingual}
\end{table}

To understand the tradeoff between multilingual coverage and English-specific performance, we evaluate our multilingual models on the same English-only benchmarks used for the English R2 models. Table \ref{tab:english-vs-multilingual} shows the comparison, with the English models performing better on English Retrieval, while the multilingual models showing stronger performance on Code and Long-document retrieval.

\subsection{Embedding Speed}
\label{sec:time}

Text embedding models are fundamental to information retrieval systems and Retrieval-Augmented Generation (RAG) applications. Organizations typically process millions of documents, with frequent updates and new content requiring continuous ingestion. This makes encoding speed as important as accuracy---a slow model can become a significant bottleneck in large-scale deployments.

\begin{table}[ht]
  \centering
  \setlength{\tabcolsep}{4pt}
  \small
  \begin{tabular}{lrrrr}
    \toprule
        Model                               &  Parameters  &  Embedding  & Encoding Speed  & Rel to Granite \\
         &  (M) $\downarrow$ & Size   & (Docs/s) $\uparrow$  &  R2 equivalent\\
        \midrule
    {\footnotesize F2LLM-v2-80M}                       &  80 & 320 & 2190 &  86.4\% \\
    \small{multilingual-e5-small}                 &  96 & 384 & 2604 & 118.9\% \\
    \rowcolor{blue!10}
    \small{granite-embedding-97m-multilingual-r2}  &  97 & 384 & 2534 & 100.0\% \\
    \small{granite-embedding-107m-multilingual}    & 107 & 384 & 3113 & 122.9\% \\
    \small{jina-embeddings-v5-text-nano}            & 239 & 768 &  307 &  16.8\% \\
    \small{harrier-oss-v1-270m}                  & 270 & 640 & 1938 & 106.0\% \\
    \small{multilingual-e5-base}                  & 278 & 768 & 2025 & 110.8\% \\
    \small{granite-embedding-278m-multilingual}    & 278 & 768 & 2164 & 118.4\% \\
    \small{embeddinggemma-300m}                     & 300 & 768 & 1349 &  73.8\% \\
    \small{gte-multilingual-base}              & 305 & 768 & 2018 & 110.4\% \\
    \small{snowflake-arctic-embed-m-v2.0}        & 305 & 768 & 2190 & 119.8\% \\
    \rowcolor{blue!10}
    \small{granite-embedding-311m-multilingual-r2} & 311 & 768 & 1828 & 100.0\% \\
    \bottomrule
  \end{tabular}
  \caption{Encoding speed comparison for multilingual models. All evaluations are done on a single Nvidia H100 GPU, with a batch size of 512 and truncating texts to 512 tokens max (to be comparable with 512 max models). The last column represents the relative speed of the given model to the size-equivalent granite multilingual R2 model. The evaluation was run with the version 5.8.0 of HuggingFace transformers - see Appendix \ref{MB-speed} for important speed considerations.}
  \label{tab:throughput}
\end{table}

\subsection{Speed vs.\ Accuracy}

\begin{figure}[!t]
    \centering
    \includegraphics[width=1.2\linewidth]{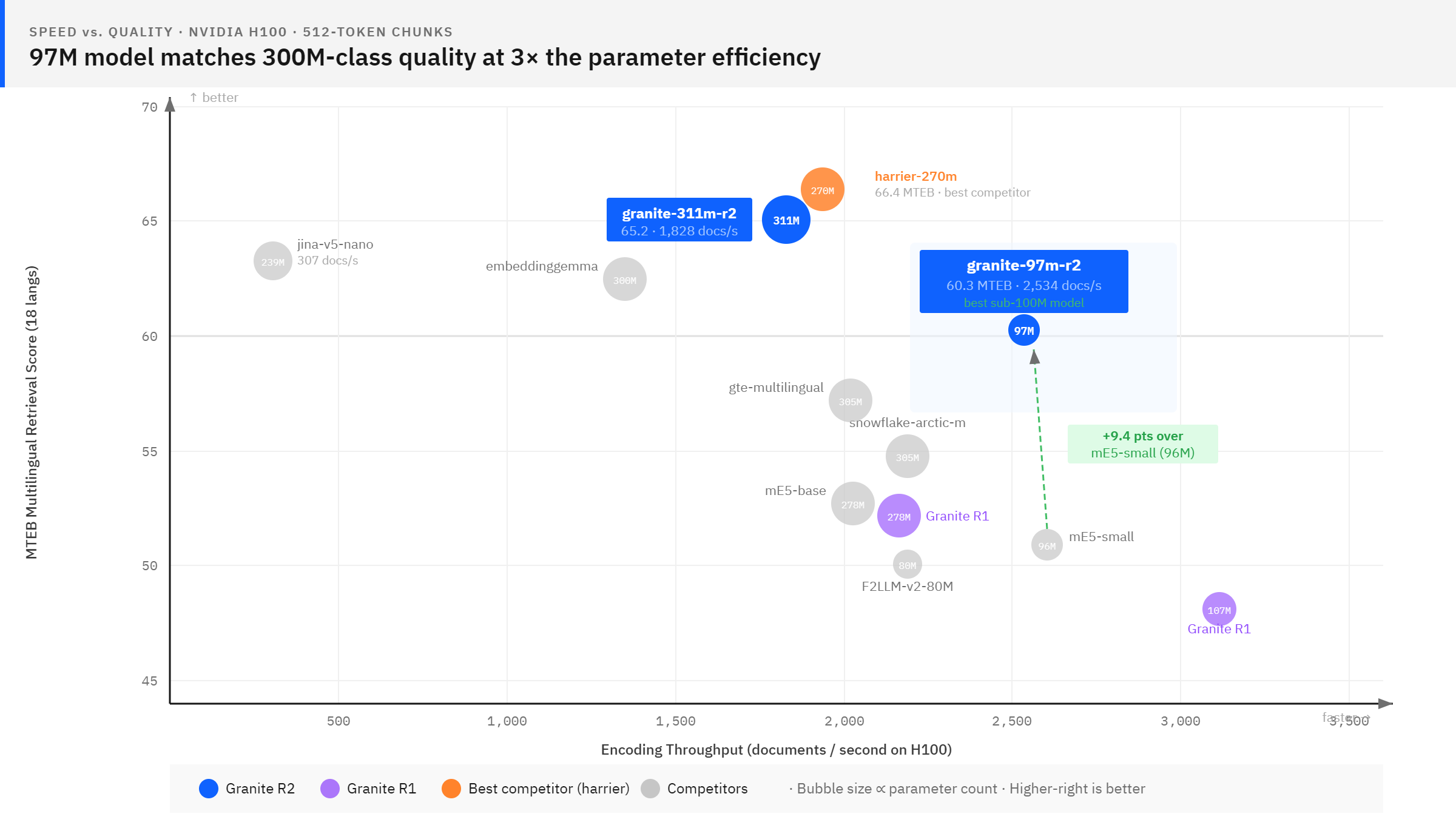}
    \caption{Speed vs.\ accuracy Pareto frontier for multilingual embedding models. The x-axis shows encoding speed (docs/sec on a single H100 GPU) and the y-axis shows MTEB-v2 Retrieval Avg. Models on or near the Pareto frontier offer the best tradeoff between throughput and retrieval quality. Granite multilingual R2 models are highlighted.}
\label{fig:speed-accuracy}
\end{figure}

We compare the two R2 multilingual models against contemporary multilingual encoders of comparable scale. Quality is reported as the MTEB Multilingual Retrieval average across 18 languages; throughput is measured in 512-token documents per second on a single NVIDIA H100 with batch size and sequence length held constant across models. Results are summarised in Figure~\ref{fig:speed-accuracy} and Table~\ref{tab:throughput}.

At the 300M-parameter tier, \texttt{granite-embedding-311m-multilingual-r2} reaches 65.2 MTEB at 1{,}828 docs/s. Among the surveyed models it is the second-strongest on retrieval quality, behind \texttt{harrier-oss-v1-270m} (66.4 MTEB, 1{,}938 docs/s), and ahead of \texttt{embeddinggemma-300m}, \texttt{gte-multilingual-base}, \texttt{snowflake-arctic-embed-m-v2.0}, and \texttt{multilingual-e5-base}. The R1 multilingual model at the same tier ($278$M parameters) scored substantially lower at comparable throughput, so the R2 release represents a large absolute gain in retrieval quality without a throughput regression in this size class.

At the sub-100M tier the gap to peer models is more pronounced. \texttt{granite-embedding-97m-multilingual-r2} reaches 60.3 MTEB at 2{,}534 docs/s, exceeding \texttt{multilingual-e5-small} (96M, 50.9 MTEB at 2{,}604 docs/s) by 9.4 points at near-identical throughput, and improving on \texttt{F2LLM-v2-80M} by a similar margin. We are not aware of a publicly available model below 100M parameters that approaches this level of multilingual retrieval quality. Relative to its own larger sibling, the 97M model trails by 4.9 MTEB points while delivering $\approx$1.4$\times$ the throughput; relative to \texttt{embeddinggemma-300m} it is within a few points of quality at $\approx$1.9$\times$ the throughput and roughly one third of the parameter count.

The two R2 models therefore occupy distinct points on the speed--quality frontier: the 311M variant targets settings where retrieval quality dominates and a $\sim$2{,}000 docs/s encoding budget is acceptable, while the 97M variant targets latency- or memory-constrained settings where the prior best multilingual option in this class sacrificed close to ten points of MTEB quality. Neither model dominates on both axes against all peers --- \texttt{harrier-270m} retains a small lead on quality at the 300M tier, and \texttt{multilingual-e5-small} retains a small lead on raw throughput at the sub-100M tier --- but both R2 models lie on the upper-right Pareto envelope of the configurations evaluated here.

\section{Conclusion}
In this work, we have presented the Granite Embedding Multilingual R2 Models, a family of specialized multilingual retrieval models designed to address the computational and accuracy requirements of enterprise-scale multilingual information retrieval systems. Our models support 200+ languages with enhanced support for 52 languages and programming code, a 32,768-token context window, and deliver state-of-the-art performance across diverse retrieval domains.

The proposed models incorporate several key contributions: (1) a full-size 311M-parameter multilingual embedding model that ranks in the top 3 of open multilingual models under 500M parameters on MTEB-v2 Retrieval, with Matryoshka Representation Learning for flexible embedding dimensionality; (2) a compact 97M-parameter multilingual model, built via model pruning and vocabulary selection, that supports long context and achieves the highest retrieval score of any open multilingual model under 100M parameters; and (3) comprehensive training on enterprise-appropriate data with transparent governance. We release these models under the Apache 2.0 license supporting both academic research and practical deployment scenarios.

\bibliography{main}

@inproceedings{
hu2024minicpm,
title={Mini{CPM}: Unveiling the Potential of Small Language Models with Scalable Training Strategies},
author={Shengding Hu and Yuge Tu and Xu Han and Ganqu Cui and Chaoqun He and Weilin Zhao and Xiang Long and Zhi Zheng and Yewei Fang and Yuxiang Huang and Xinrong Zhang and Zhen Leng Thai and Chongyi Wang and Yuan Yao and Chenyang Zhao and Jie Zhou and Jie Cai and Zhongwu Zhai and Ning Ding and Chao Jia and Guoyang Zeng and dahai li and Zhiyuan Liu and Maosong Sun},
booktitle={First Conference on Language Modeling},
year={2024},
url={https://openreview.net/forum?id=3X2L2TFr0f}
}

@misc{granite_4.1_llms,
  title        = {Granite 4.1 LLMs: How They’re Built},
  author       = {Yousaf Shah},
  publisher    = {Hugging Face},
  year         = {2026},
  url          = {"https://huggingface.co/blog/ibm-granite/granite-4-1"}
}

@inproceedings{gao-etal-2021-simcse,
    title = "{S}im{CSE}: Simple Contrastive Learning of Sentence Embeddings",
    author = "Gao, Tianyu  and
      Yao, Xingcheng  and
      Chen, Danqi",
    editor = "Moens, Marie-Francine  and
      Huang, Xuanjing  and
      Specia, Lucia  and
      Yih, Scott Wen-tau",
    booktitle = "Proceedings of the 2021 Conference on Empirical Methods in Natural Language Processing",
    month = nov,
    year = "2021",
    address = "Online and Punta Cana, Dominican Republic",
    publisher = "Association for Computational Linguistics",
    url = "https://aclanthology.org/2021.emnlp-main.552",
    doi = "10.18653/v1/2021.emnlp-main.552",
    pages = "6894--6910",
}

@misc{merrick2024arcticembedv1,
    title={Arctic-Embed: Scalable, Efficient, and Accurate Text Embedding Models},
    author={Luke Merrick and Danmei Xu and Gaurav Nuti and Daniel Campos},
    year={2024},
    eprint={2405.05374},
    archivePrefix={arXiv},
    primaryClass={cs.CL}
}

@misc{chen2024bgem3,
    title={BGE M3-Embedding: Multi-Lingual, Multi-Functionality, Multi-Granularity Text Embeddings Through Self-Knowledge Distillation},
    author={Jianlv Chen and Shitao Xiao and Peitian Zhang and Kun Luo and Defu Lian and Zheng Liu},
    year={2024},
    eprint={2402.03216},
    archivePrefix={arXiv},
    primaryClass={cs.CL}
}

@article{wang2022e5,
  title={Text Embeddings by Weakly-Supervised Contrastive Pre-training},
  author={Wang, Liang and Yang, Nan and Huang, Xiaolong and Jiao, Binxing and Yang, Linjun and Jiang, Daxin and Majumder, Rangan and Wei, Furu},
  journal={arXiv preprint arXiv:2212.03533},
  year={2022}
}

@misc{bge_embedding,
      title={C-Pack: Packaged Resources To Advance General Chinese Embedding}, 
      author={Shitao Xiao and Zheng Liu and Peitian Zhang and Niklas Muennighoff},
      year={2023},
      eprint={2309.07597},
      archivePrefix={arXiv},
      primaryClass={cs.CL}
}

@misc{nussbaum2024nomic,
      title={Nomic Embed: Training a Reproducible Long Context Text Embedder}, 
      author={Zach Nussbaum and John X. Morris and Brandon Duderstadt and Andriy Mulyar},
      year={2024},
      eprint={2402.01613},
      archivePrefix={arXiv},
      primaryClass={cs.CL}
}

@misc{zhang2024mgte,
  title={mGTE: Generalized Long-Context Text Representation and Reranking Models for Multilingual Text Retrieval}, 
  author={Xin Zhang and Yanzhao Zhang and Dingkun Long and Wen Xie and Ziqi Dai and Jialong Tang and Huan Lin and Baosong Yang and Pengjun Xie and Fei Huang and Meishan Zhang and Wenjie Li and Min Zhang},
  year={2024},
  eprint={2407.19669},
  archivePrefix={arXiv},
  primaryClass={cs.CL},
  url={https://arxiv.org/abs/2407.19669}, 
}

@misc{li2023gte,
  title={Towards General Text Embeddings with Multi-stage Contrastive Learning}, 
  author={Zehan Li and Xin Zhang and Yanzhao Zhang and Dingkun Long and Pengjun Xie and Meishan Zhang},
  year={2023},
  eprint={2308.03281},
  archivePrefix={arXiv},
  primaryClass={cs.CL},
  url={https://arxiv.org/abs/2308.03281}, 
}

@misc{dunn2017searchqa,
      title={SearchQA: A New Q\&A Dataset Augmented with Context from a Search Engine}, 
      author={Matthew Dunn and Levent Sagun and Mike Higgins and V. Ugur Guney and Volkan Cirik and Kyunghyun Cho},
      year={2017},
      eprint={1704.05179},
      archivePrefix={arXiv},
      primaryClass={cs.CL}
}

@misc{abdin2024phi4technicalreport,
      title={Phi-4 Technical Report}, 
      author={Marah Abdin and Jyoti Aneja and Harkirat Behl and Sébastien Bubeck and Ronen Eldan and Suriya Gunasekar and Michael Harrison and Russell J. Hewett and Mojan Javaheripi and Piero Kauffmann and James R. Lee and Yin Tat Lee and Yuanzhi Li and Weishung Liu and Caio C. T. Mendes and Anh Nguyen and Eric Price and Gustavo de Rosa and Olli Saarikivi and Adil Salim and Shital Shah and Xin Wang and Rachel Ward and Yue Wu and Dingli Yu and Cyril Zhang and Yi Zhang},
      year={2024},
      eprint={2412.08905},
      archivePrefix={arXiv},
      primaryClass={cs.CL},
      url={https://arxiv.org/abs/2412.08905}, 
}

@misc{liu2026ministral3,
      title={Ministral 3}, 
      author={Alexander H. Liu and Kartik Khandelwal and Sandeep Subramanian and Victor Jouault and Abhinav Rastogi and Adrien Sadé and Alan Jeffares and Albert Jiang and Alexandre Cahill and Alexandre Gavaudan and Alexandre Sablayrolles and Amélie Héliou and Amos You and Andy Ehrenberg and Andy Lo and Anton Eliseev and Antonia Calvi and Avinash Sooriyarachchi and Baptiste Bout and Baptiste Rozière and Baudouin De Monicault and Clémence Lanfranchi and Corentin Barreau and Cyprien Courtot and Daniele Grattarola and Darius Dabert and Diego de las Casas and Elliot Chane-Sane and Faruk Ahmed and Gabrielle Berrada and Gaëtan Ecrepont and Gauthier Guinet and Georgii Novikov and Guillaume Kunsch and Guillaume Lample and Guillaume Martin and Gunshi Gupta and Jan Ludziejewski and Jason Rute and Joachim Studnia and Jonas Amar and Joséphine Delas and Josselin Somerville Roberts and Karmesh Yadav and Khyathi Chandu and Kush Jain and Laurence Aitchison and Laurent Fainsin and Léonard Blier and Lingxiao Zhao and Louis Martin and Lucile Saulnier and Luyu Gao and Maarten Buyl and Margaret Jennings and Marie Pellat and Mark Prins and Mathieu Poirée and Mathilde Guillaumin and Matthieu Dinot and Matthieu Futeral and Maxime Darrin and Maximilian Augustin and Mia Chiquier and Michel Schimpf and Nathan Grinsztajn and Neha Gupta and Nikhil Raghuraman and Olivier Bousquet and Olivier Duchenne and Patricia Wang and Patrick von Platen and Paul Jacob and Paul Wambergue and Paula Kurylowicz and Pavankumar Reddy Muddireddy and Philomène Chagniot and Pierre Stock and Pravesh Agrawal and Quentin Torroba and Romain Sauvestre and Roman Soletskyi and Rupert Menneer and Sagar Vaze and Samuel Barry and Sanchit Gandhi and Siddhant Waghjale and Siddharth Gandhi and Soham Ghosh and Srijan Mishra and Sumukh Aithal and Szymon Antoniak and Teven Le Scao and Théo Cachet and Theo Simon Sorg and Thibaut Lavril and Thiziri Nait Saada and Thomas Chabal and Thomas Foubert and Thomas Robert and Thomas Wang and Tim Lawson and Tom Bewley and Tom Bewley and Tom Edwards and Umar Jamil and Umberto Tomasini and Valeriia Nemychnikova and Van Phung and Vincent Maladière and Virgile Richard and Wassim Bouaziz and Wen-Ding Li and William Marshall and Xinghui Li and Xinyu Yang and Yassine El Ouahidi and Yihan Wang and Yunhao Tang and Zaccharie Ramzi},
      year={2026},
      eprint={2601.08584},
      archivePrefix={arXiv},
      primaryClass={cs.CL},
      url={https://arxiv.org/abs/2601.08584}, 
}

@inproceedings{hinton2014distilling,
  title={{Distilling the Knowledge in a Neural Network}},
  author={Geoffrey Hinton and Oriol Vinyals and Jeff Dean},
  booktitle={NeurIPS Deep Learning Worksop},
  year={2014},
}

@misc{li2024coircomprehensivebenchmarkcode,
      title={CoIR: A Comprehensive Benchmark for Code Information Retrieval Models}, 
      author={Xiangyang Li and Kuicai Dong and Yi Quan Lee and Wei Xia and Yichun Yin and Hao Zhang and Yong Liu and Yasheng Wang and Ruiming Tang},
      year={2024},
      eprint={2407.02883},
      archivePrefix={arXiv},
      primaryClass={cs.IR},
      url={https://arxiv.org/abs/2407.02883}, 
}

@misc{wang2024multilinguale5,
    title={Multilingual E5 Text Embeddings: A Technical Report},
    author={Liang Wang and Nan Yang and Xiaolong Huang and Linjun Yang and Rangan Majumder and Furu Wei},
    year={2024},
    eprint={2402.05672},
    archivePrefix={arXiv},
    primaryClass={cs.CL}
}

@misc{lee2024nvembed,
    title={NV-Embed: Improved Techniques for Training LLMs as Generalist Embedding Models},
    author={Chankyu Lee and Rajarshi Roy and Mengyao Xu and Jonathan Raiman and Mohammad Shoeybi and Bryan Catanzaro and Wei Ping},
    year={2024},
    eprint={2405.17428},
    archivePrefix={arXiv},
    primaryClass={cs.CL}
}

@misc{wang2023e5mistral,
    title={Improving Text Embeddings with Large Language Models},
    author={Liang Wang and Nan Yang and Xiaolong Huang and Linjun Yang and Rangan Majumder and Furu Wei},
    year={2023},
    eprint={2401.00368},
    archivePrefix={arXiv},
    primaryClass={cs.CL}
}

@misc{jiang2023mistral7b,
      title={Mistral 7B}, 
      author={Albert Q. Jiang and Alexandre Sablayrolles and Arthur Mensch and Chris Bamford and Devendra Singh Chaplot and Diego de las Casas and Florian Bressand and Gianna Lengyel and Guillaume Lample and Lucile Saulnier and Lélio Renard Lavaud and Marie-Anne Lachaux and Pierre Stock and Teven Le Scao and Thibaut Lavril and Thomas Wang and Timothée Lacroix and William El Sayed},
      year={2023},
      eprint={2310.06825},
      archivePrefix={arXiv},
      primaryClass={cs.CL},
      url={https://arxiv.org/abs/2310.06825}, 
}

@inproceedings{10.1145/3269206.3271800,
author = {Zamani, Hamed and Dehghani, Mostafa and Croft, W. Bruce and Learned-Miller, Erik and Kamps, Jaap},
title = {From Neural Re-Ranking to Neural Ranking: Learning a Sparse Representation for Inverted Indexing},
year = {2018},
isbn = {9781450360142},
publisher = {Association for Computing Machinery},
address = {New York, NY, USA},
url = {https://doi.org/10.1145/3269206.3271800},
doi = {10.1145/3269206.3271800},
booktitle = {Proceedings of the 27th ACM International Conference on Information and Knowledge Management},
pages = {497–506},
numpages = {10},
location = {Torino, Italy},
series = {CIKM '18}
}

@misc{xiong2020approximatenearestneighbornegative,
      title={Approximate Nearest Neighbor Negative Contrastive Learning for Dense Text Retrieval}, 
      author={Lee Xiong and Chenyan Xiong and Ye Li and Kwok-Fung Tang and Jialin Liu and Paul Bennett and Junaid Ahmed and Arnold Overwijk},
      year={2020},
      eprint={2007.00808},
      archivePrefix={arXiv},
      primaryClass={cs.IR},
      url={https://arxiv.org/abs/2007.00808}, 
}

@misc{neelakantan2022textcodeembeddingscontrastive,
      title={Text and Code Embeddings by Contrastive Pre-Training}, 
      author={Arvind Neelakantan and Tao Xu and Raul Puri and Alec Radford and Jesse Michael Han and Jerry Tworek and Qiming Yuan and Nikolas Tezak and Jong Wook Kim and Chris Hallacy and Johannes Heidecke and Pranav Shyam and Boris Power and Tyna Eloundou Nekoul and Girish Sastry and Gretchen Krueger and David Schnurr and Felipe Petroski Such and Kenny Hsu and Madeleine Thompson and Tabarak Khan and Toki Sherbakov and Joanne Jang and Peter Welinder and Lilian Weng},
      year={2022},
      eprint={2201.10005},
      archivePrefix={arXiv},
      primaryClass={cs.CL},
      url={https://arxiv.org/abs/2201.10005}, 
}

@misc{zhao2020spartaefficientopendomainquestion,
      title={SPARTA: Efficient Open-Domain Question Answering via Sparse Transformer Matching Retrieval}, 
      author={Tiancheng Zhao and Xiaopeng Lu and Kyusong Lee},
      year={2020},
      eprint={2009.13013},
      archivePrefix={arXiv},
      primaryClass={cs.CL},
      url={https://arxiv.org/abs/2009.13013}, 
}

@article{angelov20,
  author       = {Dimo Angelov},
  title        = {Top2Vec: Distributed Representations of Topics},
  journal      = {CoRR},
  volume       = {abs/2008.09470},
  year         = {2020},
  url          = {https://arxiv.org/abs/2008.09470},
  eprinttype    = {arXiv},
  eprint       = {2008.09470},
  bibsource    = {dblp computer science bibliography, https://dblp.org}
}

@inproceedings{sun2019fine,
  title={How to fine-tune bert for text classification?},
  author={Sun, Chi and Qiu, Xipeng and Xu, Yige and Huang, Xuanjing},
  booktitle={China national conference on Chinese computational linguistics},
  pages={194--206},
  year={2019},
  organization={Springer}
}

@misc{warner2024modernbert,
    title={Smarter, Better, Faster, Longer: A Modern Bidirectional Encoder for Fast, Memory Efficient, and Long Context Finetuning and Inference},
    author={Benjamin Warner and Antoine Chaffin and Benjamin Clavié and Orion Weller and Oskar Hallström and Said Taghadouini and Alexis Gallagher and Raja Biswas and Faisal Ladhak and Tom Aarsen and Nathan Cooper and Griffin Adams and Jeremy Howard and Iacopo Poli},
    year={2024},
    eprint={2412.13663},
    archivePrefix={arXiv},
    primaryClass={cs.CL}
}

@misc{gohari2025gneissweb,
    title={GneissWeb: Preparing High Quality Data for LLMs at Scale},
    author={Hajar Emami Gohari and Swanand Ravindra Kadhe and Syed Yousaf Shah and Constantin Adam and Abdulhamid Adebayo and Praneet Adusumilli and Farhan Ahmed and Nathalie Baracaldo Angel and Santosh Subhashrao Borse and Yuan-Chi Chang and Xuan-Hong Dang and Nirmit Desai and Revital Eres and Ran Iwamoto and Alexei Karve and Yan Koyfman and Wei-Han Lee and Changchang Liu and Boris Lublinsky and Takuyo Ohko and Pablo Pesce and Maroun Touma and Shiqiang Wang and Shalisha Witherspoon and Herbert Woisetschläger and David Wood and Kun-Lung Wu and Issei Yoshida and Syed Zawad and Petros Zerfos and Yi Zhou and Bishwaranjan Bhattacharjee},
    year={2025},
    eprint={2502.14907},
    archivePrefix={arXiv},
    primaryClass={cs.CL}
}

@misc{awasthy2025granite,
    title={Granite Embedding Models},
    author={Parul Awasthy and Aashka Trivedi and Yulong Li and Mihaela Bornea and David Cox and Abraham Daniels and Martin Franz and Gabe Goodhart and Bhavani Iyer and Vishwajeet Kumar and Luis Lastras and Scott McCarley and Rudra Murthy and Vignesh P and Sara Rosenthal and Salim Roukos and Jaydeep Sen and Sukriti Sharma and Avirup Sil and Kate Soule and Arafat Sultan and Radu Florian},
    year={2025},
    eprint={2502.20204},
    archivePrefix={arXiv},
    url = {https://arxiv.org/abs/2502.20204},
    primaryClass={cs.IR}
}

@misc{dao2023flashattention2,
    title={FlashAttention-2: Faster Attention with Better Parallelism and Work Partitioning},
    author={Tri Dao},
    year={2023},
    eprint={2307.08691},
    archivePrefix={arXiv},
    primaryClass={cs.LG}
}

@misc{mishra2024granite,
    title={Granite Code Models: A Family of Open Foundation Models for Code Intelligence},
    author={Mayank Mishra and Matt Stallone and Gaoyuan Zhang and Yikang Shen and Aditya Prasad and Adriana Meza Soria and Michele Merler and Parameswaran Selvam and Saptha Surendran and Shivdeep Singh and Manish Sethi and Xuan-Hong Dang and Pengyuan Li and Kun-Lung Wu and Syed Zawad and Andrew Coleman and Matthew White and Mark Lewis and Raju Pavuluri and Yan Koyfman and Boris Lublinsky and Maximilien de Bayser and Ibrahim Abdelaziz and Kinjal Basu and Mayank Agarwal and Yi Zhou and Chris Johnson and Aanchal Goyal and Hima Patel and Yousaf Shah and Petros Zerfos and Heiko Ludwig and Asim Munawar and Maxwell Crouse and Pavan Kapanipathi and Shweta Salaria and Bob Calio and Sophia Wen and Seetharami Seelam and Brian Belgodere and Carlos Fonseca and Amith Singhee and Nirmit Desai and David D. Cox and Ruchir Puri and Rameswar Panda},
    year={2024},
    eprint={2405.04324},
    archivePrefix={arXiv},
    primaryClass={cs.AI}
}

@misc{zhu2024longembed,
    title={LongEmbed: Extending Embedding Models for Long Context Retrieval},
    author={Dawei Zhu and Liang Wang and Nan Yang and Yifan Song and Wenhao Wu and Furu Wei and Sujian Li},
    year={2024},
    eprint={2404.12096},
    archivePrefix={arXiv},
    primaryClass={cs.CL}
}

@article{enevoldsen2025mmtebmassivemultilingualtext,
  author = {Kenneth Enevoldsen and Isaac Chung and Imene Kerboua and Márton Kardos and Ashwin Mathur and David Stap and Jay Gala and Wissam Siblini and Dominik Krzemiński and Genta Indra Winata and Saba Sturua and Saiteja Utpala and Mathieu Ciancone and Marion Schaeffer and Gabriel Sequeira and Diganta Misra and Shreeya Dhakal and Jonathan Rystrøm and Roman Solomatin and Ömer Çağatan and Akash Kundu and Martin Bernstorff and Shitao Xiao and Akshita Sukhlecha and Bhavish Pahwa and Rafał Poświata and Kranthi Kiran GV and Shawon Ashraf and Daniel Auras and Björn Plüster and Jan Philipp Harries and Loïc Magne and Isabelle Mohr and Mariya Hendriksen and Dawei Zhu and Hippolyte Gisserot-Boukhlef and Tom Aarsen and Jan Kostkan and Konrad Wojtasik and Taemin Lee and Marek Šuppa and Crystina Zhang and Roberta Rocca and Mohammed Hamdy and Andrianos Michail and John Yang and Manuel Faysse and Aleksei Vatolin and Nandan Thakur and Manan Dey and Dipam Vasani and Pranjal Chitale and Simone Tedeschi and Nguyen Tai and Artem Snegirev and Michael Günther and Mengzhou Xia and Weijia Shi and Xing Han Lù and Jordan Clive and Gayatri Krishnakumar and Anna Maksimova and Silvan Wehrli and Maria Tikhonova and Henil Panchal and Aleksandr Abramov and Malte Ostendorff and Zheng Liu and Simon Clematide and Lester James Miranda and Alena Fenogenova and Guangyu Song and Ruqiya Bin Safi and Wen-Ding Li and Alessia Borghini and Federico Cassano and Hongjin Su and Jimmy Lin and Howard Yen and Lasse Hansen and Sara Hooker and Chenghao Xiao and Vaibhav Adlakha and Orion Weller and Siva Reddy and Niklas Muennighoff},
  doi = {10.48550/arXiv.2502.13595},
  journal = {arXiv preprint arXiv:2502.13595},
  publisher = {arXiv},
  title = {MMTEB: Massive Multilingual Text Embedding Benchmark},
  url = {https://arxiv.org/abs/2502.13595},
  year = {2025},
}

@misc{wortsman2023stable,
    title={Stable and low-precision training for large-scale vision-language models},
    author={Mitchell Wortsman and Tim Dettmers and Luke Zettlemoyer and Ari Morcos and Ali Farhadi and Ludwig Schmidt},
    year={2023},
    eprint={2304.13013},
    archivePrefix={arXiv},
    primaryClass={cs.LG}
}

@misc{yu2024arcticembedv2,
    title={Arctic-Embed 2.0: Multilingual Retrieval Without Compromise},
    author={Puxuan Yu and Luke Merrick and Gaurav Nuti and Daniel Campos},
    year={2024},
    eprint={2412.04506},
    archivePrefix={arXiv},
    primaryClass={cs.CL}
}

@misc{awasthy2025graniteembeddingr2models,
      title={Granite Embedding R2 Models}, 
      author={Parul Awasthy and Aashka Trivedi and Yulong Li and Meet Doshi and Riyaz Bhat and Vignesh P and Vishwajeet Kumar and Yushu Yang and Bhavani Iyer and Abraham Daniels and Rudra Murthy and Ken Barker and Martin Franz and Madison Lee and Todd Ward and Salim Roukos and David Cox and Luis Lastras and Jaydeep Sen and Radu Florian},
      year={2025},
      eprint={2508.21085},
      archivePrefix={arXiv},
      primaryClass={cs.CL},
      url={https://arxiv.org/abs/2508.21085}, 
}

@inproceedings{goddard-etal-2024-arcees,
    title = "Arcee{'}s {M}erge{K}it: A Toolkit for Merging Large Language Models",
    author = "Goddard, Charles  and
      Siriwardhana, Shamane  and
      Ehghaghi, Malikeh  and
      Meyers, Luke  and
      Karpukhin, Vladimir  and
      Benedict, Brian  and
      McQuade, Mark  and
      Solawetz, Jacob",
    editor = "Dernoncourt, Franck  and
      Preo{\c{t}}iuc-Pietro, Daniel  and
      Shimorina, Anastasia",
    booktitle = "Proceedings of the 2024 Conference on Empirical Methods in Natural Language Processing: Industry Track",
    month = nov,
    year = "2024",
    address = "Miami, Florida, US",
    publisher = "Association for Computational Linguistics",
    url = "https://aclanthology.org/2024.emnlp-industry.36",
    doi = "10.18653/v1/2024.emnlp-industry.36",
    pages = "477--485",
}

@article{qwen3embedding,
  title={Qwen3 Embedding: Advancing Text Embedding and Reranking Through Foundation Models},
  author={Zhang, Yanzhao and Li, Mingxin and Long, Dingkun and Zhang, Xin and Lin, Huan and Yang, Baosong and Xie, Pengjun and Yang, An and Liu, Dayiheng and Lin, Junyang and Huang, Fei and Zhou, Jingren},
  journal={arXiv preprint arXiv:2506.05176},
  year={2025}
}

@misc{akram2026jina,
      title={jina-embeddings-v5-text: Task-Targeted Embedding Distillation}, 
      author={Mohammad Kalim Akram and Saba Sturua and Nastia Havriushenko and Quentin Herreros and Michael Günther and Maximilian Werk and Han Xiao},
      year={2026},
      eprint={2602.15547},
      archivePrefix={arXiv},
      primaryClass={cs.CL},
      url={https://arxiv.org/abs/2602.15547}, 
}

@article{wang2024multilingual,
  title={Multilingual E5 Text Embeddings: A Technical Report},
  author={Wang, Liang and Yang, Nan and Huang, Xiaolong and Yang, Linjun and Majumder, Rangan and Wei, Furu},
  journal={arXiv preprint arXiv:2402.05672},
  year={2024}
}

@misc{shao2025reasonirtrainingretrieversreasoning,
      title={ReasonIR: Training Retrievers for Reasoning Tasks}, 
      author={Rulin Shao and Rui Qiao and Varsha Kishore and Niklas Muennighoff and Xi Victoria Lin and Daniela Rus and Bryan Kian Hsiang Low and Sewon Min and Wen-tau Yih and Pang Wei Koh and Luke Zettlemoyer},
      year={2025},
      eprint={2504.20595},
      archivePrefix={arXiv},
      primaryClass={cs.AI},
      url={https://arxiv.org/abs/2504.20595}, 
}

@misc{penedo2025fineweb2pipelinescale,
  title={FineWeb2: One Pipeline to Scale Them All -- Adapting Pre-Training Data Processing to Every Language}, 
  author={Guilherme Penedo and Hynek Kydlíček and Vinko Sabolčec and Bettina Messmer and Negar Foroutan and Amir Hossein Kargaran and Colin Raffel and Martin Jaggi and Leandro Von Werra and Thomas Wolf},
  year={2025},
  eprint={2506.20920},
  archivePrefix={arXiv},
  primaryClass={cs.CL},
  url={https://arxiv.org/abs/2506.20920}, 
}

@misc{lozhkov2024fineweb-edu,
    author       = { Lozhkov, Anton and Ben Allal, Loubna and von Werra, Leandro and Wolf, Thomas },  
    title        = { FineWeb-Edu: the Finest Collection of Educational Content }, 
    year         = 2024,  
    url          = { https://huggingface.co/datasets/HuggingFaceFW/fineweb-edu },  
    doi          = { 10.57967/hf/2497 },
    publisher    = { Hugging Face }
}

@misc{gemmateam2025gemma3technicalreport,
      title={Gemma 3 Technical Report}, 
      author={Gemma Team and Aishwarya Kamath and Johan Ferret and Shreya Pathak and Nino Vieillard and Ramona Merhej and Sarah Perrin and Tatiana Matejovicova and Alexandre Ramé and Morgane Rivière and Louis Rouillard and Thomas Mesnard and Geoffrey Cideron and Jean-bastien Grill and Sabela Ramos and Edouard Yvinec and Michelle Casbon and Etienne Pot and Ivo Penchev and Gaël Liu and Francesco Visin and Kathleen Kenealy and Lucas Beyer and Xiaohai Zhai and Anton Tsitsulin and Robert Busa-Fekete and Alex Feng and Noveen Sachdeva and Benjamin Coleman and Yi Gao and Basil Mustafa and Iain Barr and Emilio Parisotto and David Tian and Matan Eyal and Colin Cherry and Jan-Thorsten Peter and Danila Sinopalnikov and Surya Bhupatiraju and Rishabh Agarwal and Mehran Kazemi and Dan Malkin and Ravin Kumar and David Vilar and Idan Brusilovsky and Jiaming Luo and Andreas Steiner and Abe Friesen and Abhanshu Sharma and Abheesht Sharma and Adi Mayrav Gilady and Adrian Goedeckemeyer and Alaa Saade and Alex Feng and Alexander Kolesnikov and Alexei Bendebury and Alvin Abdagic and Amit Vadi and András György and André Susano Pinto and Anil Das and Ankur Bapna and Antoine Miech and Antoine Yang and Antonia Paterson and Ashish Shenoy and Ayan Chakrabarti and Bilal Piot and Bo Wu and Bobak Shahriari and Bryce Petrini and Charlie Chen and Charline Le Lan and Christopher A. Choquette-Choo and CJ Carey and Cormac Brick and Daniel Deutsch and Danielle Eisenbud and Dee Cattle and Derek Cheng and Dimitris Paparas and Divyashree Shivakumar Sreepathihalli and Doug Reid and Dustin Tran and Dustin Zelle and Eric Noland and Erwin Huizenga and Eugene Kharitonov and Frederick Liu and Gagik Amirkhanyan and Glenn Cameron and Hadi Hashemi and Hanna Klimczak-Plucińska and Harman Singh and Harsh Mehta and Harshal Tushar Lehri and Hussein Hazimeh and Ian Ballantyne and Idan Szpektor and Ivan Nardini and Jean Pouget-Abadie and Jetha Chan and Joe Stanton and John Wieting and Jonathan Lai and Jordi Orbay and Joseph Fernandez and Josh Newlan and Ju-yeong Ji and Jyotinder Singh and Kat Black and Kathy Yu and Kevin Hui and Kiran Vodrahalli and Klaus Greff and Linhai Qiu and Marcella Valentine and Marina Coelho and Marvin Ritter and Matt Hoffman and Matthew Watson and Mayank Chaturvedi and Michael Moynihan and Min Ma and Nabila Babar and Natasha Noy and Nathan Byrd and Nick Roy and Nikola Momchev and Nilay Chauhan and Noveen Sachdeva and Oskar Bunyan and Pankil Botarda and Paul Caron and Paul Kishan Rubenstein and Phil Culliton and Philipp Schmid and Pier Giuseppe Sessa and Pingmei Xu and Piotr Stanczyk and Pouya Tafti and Rakesh Shivanna and Renjie Wu and Renke Pan and Reza Rokni and Rob Willoughby and Rohith Vallu and Ryan Mullins and Sammy Jerome and Sara Smoot and Sertan Girgin and Shariq Iqbal and Shashir Reddy and Shruti Sheth and Siim Põder and Sijal Bhatnagar and Sindhu Raghuram Panyam and Sivan Eiger and Susan Zhang and Tianqi Liu and Trevor Yacovone and Tyler Liechty and Uday Kalra and Utku Evci and Vedant Misra and Vincent Roseberry and Vlad Feinberg and Vlad Kolesnikov and Woohyun Han and Woosuk Kwon and Xi Chen and Yinlam Chow and Yuvein Zhu and Zichuan Wei and Zoltan Egyed and Victor Cotruta and Minh Giang and Phoebe Kirk and Anand Rao and Kat Black and Nabila Babar and Jessica Lo and Erica Moreira and Luiz Gustavo Martins and Omar Sanseviero and Lucas Gonzalez and Zach Gleicher and Tris Warkentin and Vahab Mirrokni and Evan Senter and Eli Collins and Joelle Barral and Zoubin Ghahramani and Raia Hadsell and Yossi Matias and D. Sculley and Slav Petrov and Noah Fiedel and Noam Shazeer and Oriol Vinyals and Jeff Dean and Demis Hassabis and Koray Kavukcuoglu and Clement Farabet and Elena Buchatskaya and Jean-Baptiste Alayrac and Rohan Anil and Dmitry and Lepikhin and Sebastian Borgeaud and Olivier Bachem and Armand Joulin and Alek Andreev and Cassidy Hardin and Robert Dadashi and Léonard Hussenot},
      year={2025},
      eprint={2503.19786},
      archivePrefix={arXiv},
      primaryClass={cs.CL},
      url={https://arxiv.org/abs/2503.19786}, 
}

@misc{marone2025mmbertmodernmultilingualencoder,
      title={mmBERT: A Modern Multilingual Encoder with Annealed Language Learning}, 
      author={Marc Marone and Orion Weller and William Fleshman and Eugene Yang and Dawn Lawrie and Benjamin Van Durme},
      year={2025},
      eprint={2509.06888},
      archivePrefix={arXiv},
      primaryClass={cs.CL},
      url={https://arxiv.org/abs/2509.06888}, 
}

@misc{kusupati2024matryoshkarepresentationlearning,
      title={Matryoshka Representation Learning}, 
      author={Aditya Kusupati and Gantavya Bhatt and Aniket Rege and Matthew Wallingford and Aditya Sinha and Vivek Ramanujan and William Howard-Snyder and Kaifeng Chen and Sham Kakade and Prateek Jain and Ali Farhadi},
      year={2024},
      eprint={2205.13147},
      archivePrefix={arXiv},
      primaryClass={cs.LG},
      url={https://arxiv.org/abs/2205.13147}, 
}

@article{xiao2024rar,
  author = {Xiao, Chenghao and Hudson, G Thomas and Al Moubayed, Noura},
  journal = {arXiv preprint arXiv:2404.06347},
  title = {RAR-b: Reasoning as Retrieval Benchmark},
  year = {2024},
}

@misc{vera2025embeddinggemmapowerfullightweighttext,
      title={EmbeddingGemma: Powerful and Lightweight Text Representations}, 
      author={Henrique Schechter Vera and Sahil Dua and Biao Zhang and Daniel Salz and Ryan Mullins and Sindhu Raghuram Panyam and Sara Smoot and Iftekhar Naim and Joe Zou and Feiyang Chen and Daniel Cer and Alice Lisak and Min Choi and Lucas Gonzalez and Omar Sanseviero and Glenn Cameron and Ian Ballantyne and Kat Black and Kaifeng Chen and Weiyi Wang and Zhe Li and Gus Martins and Jinhyuk Lee and Mark Sherwood and Juyeong Ji and Renjie Wu and Jingxiao Zheng and Jyotinder Singh and Abheesht Sharma and Divyashree Sreepathihalli and Aashi Jain and Adham Elarabawy and AJ Co and Andreas Doumanoglou and Babak Samari and Ben Hora and Brian Potetz and Dahun Kim and Enrique Alfonseca and Fedor Moiseev and Feng Han and Frank Palma Gomez and Gustavo Hernández Ábrego and Hesen Zhang and Hui Hui and Jay Han and Karan Gill and Ke Chen and Koert Chen and Madhuri Shanbhogue and Michael Boratko and Paul Suganthan and Sai Meher Karthik Duddu and Sandeep Mariserla and Setareh Ariafar and Shanfeng Zhang and Shijie Zhang and Simon Baumgartner and Sonam Goenka and Steve Qiu and Tanmaya Dabral and Trevor Walker and Vikram Rao and Waleed Khawaja and Wenlei Zhou and Xiaoqi Ren and Ye Xia and Yichang Chen and Yi-Ting Chen and Zhe Dong and Zhongli Ding and Francesco Visin and Gaël Liu and Jiageng Zhang and Kathleen Kenealy and Michelle Casbon and Ravin Kumar and Thomas Mesnard and Zach Gleicher and Cormac Brick and Olivier Lacombe and Adam Roberts and Qin Yin and Yunhsuan Sung and Raphael Hoffmann and Tris Warkentin and Armand Joulin and Tom Duerig and Mojtaba Seyedhosseini},
      year={2025},
      eprint={2509.20354},
      archivePrefix={arXiv},
      primaryClass={cs.CL},
      url={https://arxiv.org/abs/2509.20354}, 
}
\bibliographystyle{iclr2025_conference}

\newpage
\appendix

\section{Contributions}
\label{sec:contributions}
The Granite R2 embedding models were truly the outcome of a successful collaboration across
geographies led by Radu Florian - with contributions from IBM Watson Research Lab (WRL) lab
and India Research Lab (IRL). Parul Awasthy was the challenge lead on the project overall, calling from WRL, with Jaydeep Sen
coordinating the work from IRL. We are very grateful for the wonderful and successful collaboration
across continents - looking forward to even better models!

\subsubsection*{Encoder Model Training}
Parul Awasthy, Aashka Trivedi, Yushu Yang

\subsubsection*{Retriever Training}
Parul Awasthy, Ken Barker, Yulong Li, Aashka Trivedi, Yushu Yang

\subsubsection*{Data and Evaluation}
Parul Awasthy, Ken Barker, Juergen Bross, Meet Doshi, Radu Florian, Martin Franz, Bhavani Iyer, Vishwajeet Kumar, Yulong Li, Vignesh P, Aashka Trivedi, Todd Ward

\subsubsection*{Product Management}
Abraham Daniels, Madison Lee

\subsubsection*{Technical Leadership}
Parul Awasthy, Radu Florian, Luis Lastras, Jaydeep Sen

\section{Supported Languages}
\label{app:languages}
The 52 enhanced-support languages are: Albanian (sq), Arabic (ar), Azerbaijani (az), Bengali (bn), Bulgarian (bg), Catalan (ca), Chinese (zh), Croatian (hr), Czech (cs), Danish (da), Dutch (nl), English (en), Estonian (et), Finnish (fi), French (fr), Georgian (ka), German (de), Greek (el), Hebrew (he), Hindi (hi), Hungarian (hu), Icelandic (is), Indonesian (id), Italian (it), Japanese (ja), Kazakh (kk), Khmer (km), Korean (ko), Latvian (lv), Lithuanian (lt), Malay (ms), Marathi (mr), Norwegian (no), Persian (fa), Polish (pl), Portuguese (pt), Romanian (ro), Russian (ru), Serbian (sr), Slovak (sk), Slovenian (sl), Spanish (es), Swahili (sw), Swedish (sv), Tagalog (tl), Telugu (te), Thai (th), Turkish (tr), Ukrainian (uk), Urdu (ur), Uzbek (uz), and Vietnamese (vi). Additionally, the models are trained on programming code (Python, Go, Java, JavaScript, PHP, Ruby, SQL, C, C++) and support cross-lingual code retrieval.

\section{Detailed Retriever Performance Evaluation}
\label{app:evals}

\subsection{Matryoshka Dimension Reduction}

\begin{table}[!ht]
    \centering
    {\renewcommand{\arraystretch}{1.3}
    \begin{tabular}{l|c|c|c}
\toprule
 Dimension & MTEB Retrieval & MTEB Code & MTEB Retrieval \\
  & En. (10) & (12) & Multi. (18) \\
\hline
 768 (full)  & 52.6 & 63.9 & 65.2 \\
512 & 52.5 & 63.8 & 65.1 \\
384 & 52.1 & 63.7 & 65.0 \\
256 & 51.6 & 63.4 & 64.7 \\
128 & 50.4 & 62.3 & 63.7 \\
\bottomrule
\end{tabular}}
    \caption{Performance of granite-embedding-311m-multilingual-r2 at different Matryoshka embedding dimensions.}
\label{tab:matryoshka-perf}
\end{table}

Granite-embedding-311m-multilingual-r2 has been trained with Matryoshka dimensions \citep{kusupati2024matryoshkarepresentationlearning}, which support truncating the original 768 embedding dimensions to smaller vectors of sizes 512, 384, 256 and 128. Truncation allows for smaller memory consumption, with a minor decrease in performance, as shown in 
Table~\ref{tab:matryoshka-perf}.

\subsection{Code Retrieval Performance}

We show per code retrieval task performance of the Granite Multilingual models in Table \ref{tab:mteb-code} evaluated at a maximum sequence length of 32K.

\begin{table}[!ht]
    \centering
    \begin{adjustbox}{width=1\textwidth}
    {\renewcommand{\arraystretch}{1.3}
    \begin{tabular}{l|c|cccccccccccc}
    \toprule
    Model                               & Avg. & Apps & COIR & CESN & CFB & CFB & CSN & CSN & CT & CT & CosQA & Stack & Syn. \\
    & &  & CSN & & MT & ST & CCR & & Contest & DL &  & OverFlow QA & Text2sql \\ \hline
    \midrule
    granite-embedding-107m-multilingual &  40.7&  4.5&  46.3& 48.1  &  23.5&  58.5&  38.3&  72.7&  43.7&  18.2& 28.3& 66.6& 39.3\\
    granite-embedding-278m-multilingual &  48.5&  6.2&  52.9&  51.3&  31.4&  67.7&  43&  76.4&  60.8&  32.1& 33.4& 78.5& 47.8\\
    \textbf{granite-embedding-97m-multilingual-r2}   &  60.4&  50.2&  75.4& 53.3 &  55.2&  75.2&  51.9&  85.3&  76.5&  34.1& 34& 82.2& 52.3\\
    \textbf{granite-embedding-311m-multilingual-r2}  &  63.9&  60.6&  77.6& 56.2 &  57&  77.2&  55.5&  86.9&  83.2&  34.4& 35.4& 86.6&  55.7\\
    
    \bottomrule
    \end{tabular}}
    \end{adjustbox}
    \caption{Code Retrieval Performance on the MTEB Code Benchmark. All scores are NDCG@10. \textit{CSN, CESN, CFB,
CT} is short for \textit{CodeSearchNet, CodeEdistSearchNet, CodeFeedBack, CodeTrans}, respectively.}
\label{tab:mteb-code}
\end{table}

\subsection{Long-Context Retrieval Performance}

\begin{table}[ht!]
\begin{adjustbox}{width=1\textwidth}
{\renewcommand{\arraystretch}{1.3}
\begin{tabular}{l|c|ccccccc}
\toprule
Model & MSL & \multicolumn{7}{c}{LongEmbed} \\
   &  & NQA & Needle & PassKey & QMSum & SFD & 2WmQA & Avg. \\
\hline
granite-embedding-107m-multilingual & 512   & 22.9 & 21.5 & 28.3 & 21.1 & 58.5 & 53.7 & 34.3 \\
granite-embedding-278m-multilingual & 512  &  23.8 & 25.8 & 38.5 & 21.8 & 60.6 & 55.8 & 37.7\\
\textbf{granite-embedding-97m-multilingual-r2} & 32768    & 46.0 & 53.3 & 82.8  &  41.2  & 90.1 & 79.4 & 65.5 \\
\textbf{granite-embedding-311m-multilingual-r2} & 32768   &  51.1 & 55.5 & 95.5  & 47.5 & 95.7 & 85.2 & 71.1 \\
\bottomrule
\end{tabular}}
\end{adjustbox}
\caption{Long Context Retrieval Performance on LongEmbed. MSL depicts the maximum sequence length of the embedding model. All scores are NDCG@10 except the Needle and Passkey subsets, which report Accuracy@1. \textit{NQA, SFD,
2WmQA} is short for \textit{NarrativeQA, SummScreenFD, 2WikiMultihopQA}, respectively.}
\label{tab:longctx}
\end{table}

Table~\ref{tab:longctx} shows the strong performance of the Granite Embedding Multilingual R2 models on the long context benchmark LongEmbed, evaluated at a maximum sequence length of 32K.

\subsection{Per-Language MIRACL Retrieval}

\begin{table}[!t]                                                                   \centering                                                                    \begin{adjustbox}{width=1.0\textwidth}
{\renewcommand{\arraystretch}{1.3}%
\begin{tabular}{l|c|c|c|c}                                                          \toprule                                                                             Language  & granite-embeddings & granite-embeddings & granite-embeddings & granite-embeddings \\                                             
& -107m-multilingual & -278m-multilingual & -97m-multilingual-r2 & -311m-multilingual-r2 \\                                        
\hline                                                                              Arabic (ar) & 63.98 & 65.74 & 63.76 & 67.93 \\                                      Bengali (bn) & 65.48 & 68.05 & 66.29 & 70.85 \\                                     Chinese (zh) & 52.02 & 54.91 & 55.07 & 58.35 \\                                     English (en) & 49.04 & 51.13 & 45.39 & 45.46 \\                                     Finnish (fi) & 66.19 & 67.92 & 65.62 & 68.47 \\                                      French (fr) & 50.45 & 51.88 & 48.72 & 52.08 \\                                     German (de) & 47.41 & 49.70 & 49.59 & 50.97 \\                                      Hindi (hi) & 44.28 & 47.72 & 47.09 & 51.77 \\                                        Indonesian (id) & 46.04 & 47.51 & 45.98 & 46.93 \\                                 Japanese (ja) & 60.87 & 63.51 & 58.37 & 62.52 \\                                    Korean (ko) & 60.70 & 60.93 & 55.21 & 59.15 \\                                       Persian (fa) & 48.76 & 51.83 & 49.09 & 52.39 \\                                    Russian (ru) & 51.43 & 54.51 & 51.30 & 56.86 \\                                     Spanish (es) & 49.71 & 51.94 & 48.64 & 51.01 \\                                     Swahili (sw) & 59.11 & 61.27 & 57.70 & 66.18 \\                                     Telugu (te) & 78.29 & 79.28 & 75.34 & 80.38 \\                                      Thai (th) & 72.03 & 73.54 & 65.21 & 71.73 \\                                        Yoruba (yo) & 64.69 & 68.70 & 61.19 & 63.57 \\  
\midrule                                                                  \textbf{Average} & 57.25 & 59.45 & 56.09 & 59.81 \\                                                                                         
      \bottomrule                                                                   \end{tabular}}                                                                \end{adjustbox}                                                               \caption{Per-Language MIRACL Retrieval Performance (nDCG@10).}                \label{tab:miracl}                                                               \end{table}    

We show the per-language performance of the Granite Multilingual models in Table~\ref{tab:miracl}.

\subsection{Runtime Speed for ModernBERT models} \label{MB-speed}

In late April 2026, we observed that our ModernBERT-based embedding
models had roughly halved in throughput, with no corresponding change in
our own training or inference code. Initial debugging explored several
plausible hypotheses --- attention backend fallback ordering (whether the
model was silently dropping from FlashAttention to SDPA or eager
attention), tokenizer maximum-length configuration affecting padding
behaviour, and normalisation-related issues --- each of which turned out
to be a dead end. The symptoms pointed in multiple directions, which
made the regression difficult to localise.

The root cause was a HuggingFace \texttt{transformers} library upgrade
from version 4.57 to 5.1. In that upgrade, ModernBERT's full-model
unpadding optimization was removed as part of a broader effort to
reduce technical debt and standardize model implementations across the
library. The optimization had allowed the model to skip computation on
padding tokens throughout the entire transformer stack, which provided
a substantial speedup for batches with variable-length sequences. The
removal was a reasonable architectural decision --- maintaining bespoke
optimization paths for individual models creates long-term maintenance
burden --- but the practical effect for downstream users is that the
model now processes padding tokens through every layer, effectively
doubling compute for typical mixed-length workloads. Table~\ref{tab:speed-pinned}
reports throughput under the pinned earlier release (4.57.6), where the
optimization is still present.

In Table~\ref{tab:speed-merged} we show the impact of changing the code from release 
4.57.6 to 5.8.0, to make the reader aware of the change. Note that there is a 
workaround, using a different collator, that not only recovers the performance drop,
but actually improves performance relative to 4.57.6, but you need (at this point)
custom code in the SentenceTransformer.encode method for it. For details, see our jupyter 
notebook here \href{https://github.com/ibm-granite/granite-embedding-models/blob/multilingual_r2_examples/code/collator_sentence_transformer.ipynb
}{Granite Embedding Collator notebook}.

This case illustrates a recurring challenge at the boundary between ML
infrastructure and upstream libraries: performance characteristics can
depend on internal implementation details that may legitimately change
as libraries evolve. A routine version upgrade can shift throughput
without any functional change in model outputs, and diagnosing the
cause requires bisecting across dependency versions rather than commits
in the model repository --- a search space that is easy to overlook
when the natural first instinct is to inspect one's own code.

\begin{table}[ht]
  \centering
  \footnotesize
  \setlength{\tabcolsep}{4pt}
  \begin{tabular}{lrrrr}
    \toprule
        Model                               &  Params  &  Emb.  & Infer. Speed  & Rel to Granite \\
         &  (M) $\downarrow$ & Size   & (spans/s) $\uparrow$  &  R2 equivalent\\
    \midrule
    \small{F2LLM-v2-80M}                          &  80 & 320 & 2645 &  80.9\% \\
    {\small multilingual-e5-small}                    &  96 & 384 & 2670 &  81.7\% \\
    \rowcolor{gray!15}
    {\small granite-embedding-97m-multilingual-r2}  &  97 & 384 & 3268 & 100.0\% \\
    {\small granite-embedding-107m-multilingual}    & 107 & 384 & 2941 &  90.0\% \\
    \midrule
    {\small jina-embeddings-v5-text-nano}               & 239 & 768 & 1115 &  37.7\% \\
    {\small harrier-oss-v1-270m}                     & 270 & 640 & 2004 &  67.7\% \\
    {\small multilingual-e5-base}                     & 278 & 768 & 2084 &  70.4\% \\
    {\small granite-embedding-278m-multilingual}    & 287 & 768 & 2117 &  71.5\% \\
    {\small embeddinggemma-300m}                        & 300 & 768 & 1256 &  42.4\% \\
    {\small gte-multilingual-base}                 & 305 & 768 & --   & --      \\
    {\small snowflake-arctic-embed-m-v2.0}           & 305 & 768 & --   & --      \\
    \rowcolor{gray!15}
    {\small granite-embedding-311m-multilingual-r2} & 311 & 768 & 2960 & 100.0\% \\
    \bottomrule
  \end{tabular}
  \caption{Encoding throughput (documents per second) on a single NVIDIA H100 with 512-token inputs, measured with Hugging Face \texttt{transformers}~4.57.6. We observed a throughput regression in version 5.1.0 affecting all evaluated models, so we report numbers under the pinned earlier release. Relative speed compares each model to its tier's Granite R2 reference: granite-97m-r2 for sub-110M models and granite-311m-r2 for 230M+ models. Dashes mark models not yet re-measured under the pinned toolkit version.}
  \label{tab:speed-pinned}
\end{table}

\begin{table}[t]
\centering
\small
\begin{tabular}{lrrrrr}
\toprule
Model & Params & Embed. & 4.57.6 & 5.8.0 & $\Delta$ \\
      & (M)    & Size   & (docs/s) & (docs/s) &  \\
\midrule
F2LLM-v2-80M                                 & 80  & 320 & 2{,}645 & 2{,}190 & $-17.2\%$ \\
multilingual-e5-small                        & 96  & 384 & 2{,}670 & 2{,}604 & $-2.5\%$  \\
\textbf{granite-embedding-97m-multilingual-r2} & \textbf{97}  & \textbf{384} & \textbf{3{,}268} & \textbf{2{,}534} & $\mathbf{-22.5\%}$ \\
granite-embedding-107m-multilingual          & 107 & 384 & 2{,}941 & 3{,}113 & $+5.8\%$  \\
\midrule
jina-embeddings-v5-text-nano                 & 239 & 768 & 1{,}115 &    307  & $-72.5\%$ \\
harrier-oss-v1-270m                          & 270 & 640 & 2{,}004 & 1{,}938 & $-3.3\%$  \\
multilingual-e5-base                         & 278 & 768 & 2{,}084 & 2{,}025 & $-2.8\%$  \\
granite-embedding-278m-multilingual          & 287 & 768 & 2{,}117 & 2{,}164 & $+2.2\%$  \\
embeddinggemma-300m                          & 300 & 768 & 1{,}256 & 1{,}349 & $+7.4\%$  \\
gte-multilingual-base                        & 305 & 768 &   ---   & 2{,}018 &   ---     \\
snowflake-arctic-embed-m-v2.0                & 305 & 768 &   ---   & 2{,}190 &   ---     \\
\textbf{granite-embedding-311m-multilingual-r2} & \textbf{311} & \textbf{768} & \textbf{2{,}960} & \textbf{1{,}828} & $\mathbf{-38.2\%}$ \\
\bottomrule
\end{tabular}
\caption{Encoding throughput (documents per second) on a single NVIDIA
H100 with batch size 512 and 512-token inputs, measured under two
HuggingFace \texttt{transformers} releases: 4.57.6 (with full-model
unpadding for ModernBERT) and 5.8.0 (after the optimization was
removed). $\Delta$ reports the relative change from 4.57.6 to 5.8.0.
The regression is concentrated in ModernBERT-based models (Granite R2,
jina-v5); XLM-RoBERTa-based competitors are within a few percent
across the two releases. Dashes mark configurations not yet re-measured
on the pinned earlier release.}
\label{tab:speed-merged}
\end{table}

\section{Context Length Scaling Analysis}
\label{app:context-scaling}

\begin{table}[!ht]
    \centering
    {\renewcommand{\arraystretch}{1.3}
    \begin{tabular}{l|ccccccc}  
    \toprule
       \multicolumn{1}{r|}{\textit{Max Seq. Len} $\rightarrow$} & 512 & 1024 & 2048 & 4096 & 8192 & 16384 & 32768 \\ 
        \hline
        gte-multilingual-base (max 8192) & 38.2	& 43.6 & 	49.7 & 	59.1 & 	62.1 & \multicolumn{2}{c}{---} \\
        granite-embedding-97m-multilingual-r2 & 39.1 & 	46.4 & 	52.9 & 	59.0 & 	62.9 & 	64.3 & 	65.5  \\
        granite-embedding-311m-multilingual-r2 &  40.3 & 47.7 &	54.1 &	61.3 &	67.1 &	70.7 &	71.7  \\
        
    \bottomrule
    \end{tabular}}
    \caption{Effect of maximum sequence length at inference on long-context retrieval performance on LongEmbed Benchmark. Models are evaluated with truncation at varying sequence lengths to show how performance scales with available context. Competitors are capped at their native max sequence length. Average scores reported across the 6 tasks of the LongEmbed benchmark}
\label{tab:context-scaling}
\end{table}

To demonstrate that the extended 32,768-token context window provides practical benefits, we evaluate retrieval performance as a function of maximum sequence length at inference time. Table \ref{tab:context-scaling} shows performance on the LongEmbed benchmark improves as the allowed context length increases.

\section{Retriever Training Hyperparameters}
\label{app:hyperparameters}

The hyperparameters used for different stages of retrieval training are indicated in Table~\ref{tab:hps}.
\begin{table}[h]
    \centering
    \setlength{\tabcolsep}{0.5em}
    \begin{adjustbox}{width=1.0\textwidth}
    {\renewcommand{\arraystretch}{1.3}
    \begin{tabular}{l|c|ccccc}
    \toprule
        Model & Stage & LR & Batch Size  & Steps & Seq. Len & Rope Theta\\
        \hline
        \multirow{3}{*}{granite-embedding-311m-multilingual-r2} & Contrastive FT & $1e-4$ &  9600 & 40000 & 512 & 10000 \\
        & Contrastive KD & $8e-5$ & 128 & 8000 & 512 & 10000 \\
        & Long-Context Extension & $8e-6$ & 32 & 10000 &  4096 & 150000 \\
        \midrule
        \multirow{3}{*}{granite-embedding-97m-multilingual-r2} & Contrastive FT &  $1e-4$ &  9600 & 40000 & 512 & 10000\\
        & Contrastive KD &  $8e-5$ & 256 & 12000 & 512 & 10000  \\
        & Long-Context Extension & $8e-6$ & 128 & 10000& 4096 & 150000  \\
    \bottomrule
    \end{tabular}}
    \end{adjustbox}
    \caption{Retriever Training Hyperparameters. Batch size refers to the global batch size, and rope theta refers to the global rope theta. FT and KD refers to finetuning and knowledge distillation respectively.}
\label{tab:hps}
\end{table}

\end{document}